\documentclass[noshowpacs,prd,aps,superscriptaddress,floatfix,twocolumn,nofootinbib]{revtex4-2}
\usepackage[normalem]{ulem}
\usepackage{amsmath}
\usepackage{graphicx}
\usepackage{hyperref}
\hypersetup{
     colorlinks=true,
     linkcolor=blue,
     filecolor=blue,
     citecolor = orange,      
   }
\usepackage[dvipsnames]{xcolor}
\usepackage{siunitx}
\def\sss{\scriptscriptstyle}
\def\Mgp{M_{\sss\rm GP}}
\def\xigp{\xi_{\sss\rm GP}}
\def\Vgp{V_{\sss\rm GP}}
\def\rhogp{\rho_{\sss\rm GP}}
\def\cgp{c_{\sss\rm GP}}

\begin{document}
\title{Quantum backreaction in an analog black hole}
\author{G. Ciliberto}
\affiliation{Laboratoire d'Informatique de Paris 6, CNRS, Sorbonne Universit\'e, 4 place Jussieu, 75005 Paris, France}
\affiliation{Universit\'e Paris-Saclay, CNRS, LPTMS, 91405, Orsay, France}
\affiliation{Physikalisches Institut, Albert-Ludwigs-Universit\"at Freiburg,
Hermann-Herder-Stra{\ss}e 3, 79104 Freiburg, Germany}
\author{R. Balbinot}
\affiliation{Dipartimento di Fisica e Astronomia dell'Universit\`a di Bologna and INFN sezione di Bologna, Via Irnerio 46, 40126 Bologna, Italy}
\author{A. Fabbri}
\affiliation{Departamento de F\'isica Te\'orica
and IFIC, Universidad de Valencia-CSIC, Calle Dr. Moliner 50, 46100 Burjassot, Spain}
\author{N. Pavloff}
\affiliation{Universit\'e Paris-Saclay, CNRS, LPTMS, 91405, Orsay, France}
\affiliation{Institut Universitaire de France}

\begin{abstract}
We extend the Gross-Pitaevskii equation to incorporate the effect of
quantum fluctuations onto the flow of a weakly interacting
Bose-Einstein condensate. Applying this framework to an analog black
hole in a quasi-one-dimensional, transonic flow, we investigate how
acoustic Hawking radiation back-reacts on the background
condensate. Our results point to the emergence of stationary density
and velocity undulations in the supersonic region (analogous to the
black hole interior) and enable to evaluate the change in upstream and
downstream Mach numbers caused by Hawking radiation.  These findings
provide new insight into the interplay between quantum fluctuations
and analog gravity in Bose-Einstein condensates.
\end{abstract}

\maketitle

As first noted in Ref. \cite{Garay2000}, a one-dimensional weakly
interacting Bose gas provides an excellent platform for realizing an
analogy originally proposed by Unruh. According to this analogy, the
interface between the supersonic and the subsonic regions of a
transonic flow acts as a ``sonic horizon'' mimicking the event horizon
of a gravitational black hole \cite{Unruh1981}.  Within the
Bose-Einstein condensate (BEC) framework, this line of research
culminated in the experimental observation of analog Hawking radiation
\cite{deNova2019}, manifested as nonlocal correlations in density
fluctuations emitted from the acoustic horizon \cite{Balbinot2008}.

From an experimental standpoint, the appeal of BEC platforms lies in
their ultralow temperatures, their paradigmatic quantum nature, and
the high degree of experimental control achievable in such systems. On
the theoretical side, the Unruh analogy arises naturally: following
Bogoliubov’s approach \cite{Bogoliubov1947}, it is customary to
separate a classical field --representing the background flow that
serves as an effective metric-- from quantum fluctuations. However, the
Bogoliubov approach is approximate, and it is natural to seek
improvements by incorporating higher-order interactions among the
classical and/or quantum components of the total field. This topic has
been extensively studied in BEC physics, beginning with the pioneering
work of Beliaev \cite{Beliaev1958a,Beliaev1958b}.

In the context of analog gravity, this line of research is referred
to as quantum backreaction, drawing the analogy with the study of how
Hawking radiation affects the black hole metric
\cite{Bardeen1981,York1984,Balbinot1986,Fabbri_2005,Hu2007}. In
General Relativity, this problem is notoriously difficult and
fundamentally constrained by the absence of a quantum theory of
gravitation (for a recent review on the backreaction problem, see
Ref. \cite{delRio2025} and references therein). In contrast, the
prospects are far more promising in BEC physics, where one effectively
has a ``theory of everything'' since a single quantum field
simultaneously governs the dynamics of both the effective metric and
its quantum fluctuations. As a result, a number of theoretical studies
addressed this issue (actually in a more general framework than the
mere analogous Hawking radiation), both numerically
\cite{Robertson2018,Chatrchyan2018,Butera2019,Butera2023} and using
perturbative methods
\cite{Balbinot2005,Balbinot2005a,Schultzhold2005,Liberati2020,Baak2022,Pal2024}.
The latter closely parallel the semiclassical approach used in General
Relativity: the equation governing the classical field (analogous to
Einstein’s equations) is modified by a contribution from the
expectation value of the quantum fluctuations, encompassed in the
equivalent of a stress-energy tensor.

 However, in one-dimensional BECs, such approaches face technical
 challenges physically rooted in the Hohenberg-Mermin-Wagner theorem:
 at this dimensionality, the standard Gross-Pitaevskii description is
 particularly simple, but quantum phase fluctuations diverge at long
 wavelengths. A way around issues of this type has been proposed by
 Popov \cite{Popov1971,Popov1972a,Popov1972b}, who developed an
 amplitude-phase formalism to treat the long-wavelength degrees of
 freedom of a Bose field operator. This approach has since proven
 highly effective in BEC physics
 \cite{Shevchenko1992,Petrov2000,Andersen2002,Mora2003,Petrov2004,Chiocchetta2013,Ji2015,Micheli2022,Duval2023,Micheli2024}.
 In the analysis below, we integrate it with the perturbative
 framework developed in Refs. \cite{Griffin1996, Shi1998,
   Giorgini1998, Fedichev1998, Zaremba1999,Giorgini2000}.  This set of
 studies incorporates corrections beyond the leading Gross–Pitaevskii
 order, capturing both the normal and anomalous averages [defined in
   Eq. \eqref{eq.anomal}]. Such contributions are crucial for a
 consistent treatment of both quantum and finite-temperature
 fluctuations.

The paper is organized as follows. In Sec. \ref{sec.I}, we introduce
the model and the symmetry-breaking approach, which involves
decomposing the quantum field operator into a classical order
parameter and a (small) quantum correction. In Sec. \ref{sec.AP}, we
present an alternative, though not identical, separation that proves
better suited to our objectives.  Section \ref{sec.perturb} details
our perturbative expansion. The leading order corresponds to the
standard Gross-Pitaevskii equation (Sec. \ref{sec.GPE}). The next
order -- the Bogoliubov level -- is presented in
Sec. \ref{sec.Pert_Bogo}, and is used in Sec. \ref{sec.averages} to
compute the source terms that appear in the final step of our
expansion. This last step, presented in Sec. \ref{sec.BRClass}, leads
to a modified equation for the classical order parameter that
incorporates the effects of quantum fluctuations. The equations
obtained in Sec. \ref{sec.BRClass} are general and apply to time- and
space-dependent configurations in arbitrary dimensions.  In
Sec. \ref{sec.Stat.BR} we focus on stationary states. We first examine
a uniform system in Sec. \ref{sec.BR.uni.statio} to benchmark our
approach, and then turn in Sec. \ref{sec.analog.stat.1D} to the
specific case of a one-dimensional analog black hole—the primary
system for which our theoretical framework has been devised.  Our
conclusions and prospects for future work are presented in
Sec. \ref{sec.conclusion}. Relevant results from earlier studies, as
well as technical aspects, are summarized in the
appendixes. Appendixes \ref{app.A} and \ref{app.modes} present the
acoustic black-hole configurations we consider and the associated
quantum (outgoing or ingoing) modes. These modes are used in Appendix
\ref{app.C} as a basis for expanding the quantum fluctuation field,
which makes it possible to explicitly compute source terms relevant to
our backreaction equations. Appendix \ref{app.B} details some steps of
the computations presented in the main text.  Our method is applied in
Sec.  \ref{sec.analog.stat.1D} for a specific type of back hole
configuration (the so-called ``waterfall''), and Appendix \ref{app.E}
complements this discussion by presenting results for different
analogue settings.

\section{Self-consistent approach}\label{sec.I}

We consider bosons of mass $m$ interacting through a point-like
potential in a $d$-dimensional space ($d=1$, 2 or 3).  The Bose
quantum field operator $\hat\Psi(\boldsymbol{x},t)$ obeys the
Heisenberg equation:
\begin{equation}\label{eq1}
  i\hbar\partial_t \hat\Psi =
  -\frac{\hbar^2}{2m}\boldsymbol{\nabla}^2\hat\Psi +
  \left[U -\mu + 
g \hat\Psi^\dagger \hat\Psi\right]\hat\Psi.
\end{equation}
In this equation $\mu$ is the chemical potential, $g$ ($>0$) is the
intensity of the point-like inter-particle repulsive potential and
$U(\boldsymbol{x})$ an external potential.

The Bogoliubov approach amounts to separate in a
Bose-condensed system a classical and a quantum contribution. This can
be achieved by writing \cite{Hohenberg1965,Anderson1966}
\begin{equation}\label{eq2}
  \hat\Psi(\boldsymbol{x},t)=\Phi(\boldsymbol{x},t)
  +\hat\psi(\boldsymbol{x},t),
\end{equation}
where
\begin{equation}\label{eq3}
  \Phi = \langle \hat{\Psi}\rangle \quad\mbox{and thus}\quad
  \langle\hat\psi\rangle=0.
\end{equation}
$\Phi(\boldsymbol{x},t)$ is known as the order parameter.  In
Eq. \eqref{eq3} the average is taken with respect to some statistical
operator which needs not correspond to thermodynamical equilibrium
(and could not in the case of an analog black hole).  The
fact that $\langle \hat{\Psi}\rangle\neq 0$ implies that this is not a
number conserving state, but rather a coherent superposition of states with
different numbers of particles.

The field $\hat\psi$ describes quantum fluctuations (in all the text
we note quantum operators acting in Fock space with hats). It obeys
--as does $\hat\Psi$-- the standard Bose commutation rules
\begin{equation}\label{eq.commu}
\begin{split}
& \left[\hat\psi(\boldsymbol{x},t),
\hat\psi^\dagger(\boldsymbol{y},t)\right]=
    \delta(\boldsymbol{x}-\boldsymbol{y}),\\
&    \left[\hat\psi(\boldsymbol{x},t),
\hat\psi(\boldsymbol{y},t)\right]=
    0,
\end{split}
\end{equation}
where $[\,,]$ denotes the commutator. 

Eq. \eqref{eq2} yields
\begin{equation}\label{eq4}
\begin{split}
  \hat\Psi^\dagger\hat\Psi = & |\Phi|^2 + \Phi \hat\psi^\dagger
  + \Phi^* \hat\psi 
    + \hat\psi^\dagger\hat\psi, \\
    \hat\Psi\hat\Psi = & \Phi^2 + 2 \Phi \hat\psi + \hat\psi\hat\psi, \\
    \hat\Psi^\dagger\hat\Psi\hat\Psi = &
    |\Phi|^2\Phi + 2 |\Phi|^2 \hat\psi + \Phi^2 \hat\psi^\dagger\\ 
    & + 2\Phi \hat\psi^\dagger\hat\psi+
    \Phi^* \hat\psi\hat\psi+\hat\psi^\dagger\hat\psi\hat\psi.
\end{split}
\end{equation}
The expectation values of expressions \eqref{eq4} read
\begin{equation}\label{eq6}
\begin{split}
&    \langle\hat\Psi^\dagger\hat\Psi\rangle =  |\Phi|^2
    +\tilde n, \quad
\langle\hat\Psi\hat\Psi\rangle = \Phi^2 + \tilde{m}, \\
& \langle \hat\Psi^\dagger\hat\Psi\hat\Psi\rangle =  |\Phi|^2\Phi 
    + 2\Phi \tilde{n}+ \Phi^* \tilde{m}+
    \langle\hat\psi^\dagger\hat\psi\hat\psi\rangle,
\end{split}
\end{equation}
where 
\begin{equation}\label{eq.anomal}
    \tilde{n}(\boldsymbol{x},t)=\langle \hat\psi^\dagger\hat\psi\rangle
    \quad\mbox{and}\quad 
\tilde{m}(\boldsymbol{x},t)=\langle \hat\psi\hat\psi\rangle
\end{equation}
are known as the normal and anomalous averages, respectively.  From these
expressions, the average of Eq. \eqref{eq1} reads \cite{Zaremba1999}
\begin{equation}\label{eq7}
\begin{split}
  i\hbar\partial_t \Phi = &
  \left[-\frac{\hbar^2}{2 m}\boldsymbol{\nabla}^2
    + U-\mu+g |\Phi|^2 + 2 g \tilde{n} \right] \Phi \\
    & + g \tilde{m} \Phi^* + g
\langle\hat\psi^\dagger\hat\psi\hat\psi\rangle .
\end{split}
\end{equation}
The difference between Eqs. \eqref{eq7} and \eqref{eq1} reads
\cite{Zaremba1999}
\begin{equation}\label{eq8}
\begin{split}
i\hbar\partial_t\hat{\psi}= &         
\left[-\frac{\hbar^2}{2 m}\boldsymbol{\nabla}^2
  + U-\mu+2 g |\Phi|^2  \right] \hat\psi
+ g \Phi^2 \hat\psi^\dagger\\
& + 2 g \Phi(\hat{\psi}^\dagger\hat\psi-\tilde{n})
+ g \Phi^*(\hat{\psi}\hat\psi-\tilde{m})\\
& +
g (\hat\psi^\dagger\hat\psi\hat\psi
-
\langle\hat\psi^\dagger\hat\psi\hat\psi\rangle)
.
\end{split}
\end{equation}
At this point  Eqs. \eqref{eq7} and
\eqref{eq8} are exact. They
form a self consistent system describing the reciprocal
effects of the classical and quantum fields, $\Phi$ and $\hat\psi$
respectively. We will solve them perturbatively, assuming that the
effects of the quantum fluctuations on the classical background are
small.

\section{Amplitude-phase formalism}\label{sec.AP}

As an alternative to the decomposition \eqref{eq2} we use an
amplitude-phase formalism which consists in writing the quantum field as
\begin{equation}\label{eq9a}
    \hat\Psi=\exp\{i(\Theta + \hat{\theta})\} \sqrt{\rho(1+\hat{\eta})},
\end{equation}
where $\Theta(\boldsymbol{x},t)$ and $\rho(\boldsymbol{x},t)$ are the
classical phase and density fields, respectively. The Hermitian field
operators $\hat\theta(\boldsymbol{x},t)$ and
$\hat\eta(\boldsymbol{x},t)$ describe quantum fluctuations of the
phase and of the relative density, respectively, with
$\langle\hat\theta\rangle=0=\langle\hat\eta\rangle$.

The perturbative expansions based on decomposition \eqref{eq2} and
\eqref{eq9a} are not equivalent.  To make the difference explicit, we
carry out a series expansion of expression \eqref{eq9a}:
\begin{equation}\label{eq9b}
    \hat\Psi=\varphi \left(1+i \hat{\theta}+ \tfrac12 \hat\eta + 
    \tfrac{i}{2} \hat{\theta}\hat{\eta}-\tfrac12\hat\theta^2
    -\tfrac18 \hat{\eta}^2+\cdots
    \right),
\end{equation}
where
\begin{equation}\label{eq10}
  \varphi(\boldsymbol{x},t)=
  \sqrt{\rho(\boldsymbol{x},t)}\exp\{i\Theta(\boldsymbol{x},t)\}.
\end{equation}
In all the following it will suffice to keep in this expansion only
the terms which have been explicitly written down in
\eqref{eq9b}. This amounts to including fluctuations up
  to quadratic order in $\hat{\eta}$ and $\hat{\theta}$ and to write
\begin{equation}\label{eq11}
    \hat\Psi=\varphi(1+\hat{a}+\hat{A})
\end{equation}
where
\begin{equation}\label{eq12}
\begin{split}
    \hat{a}(\boldsymbol{x},t)& =i \hat{\theta}+ \tfrac12 \hat\eta,
    \\
    \hat{A}(\boldsymbol{x},t) & =
    \tfrac{i}{2} \hat{\theta}\hat{\eta}-\tfrac12\hat\theta^2
    -\tfrac18 \hat{\eta}^2\\
    & =-\tfrac12 \hat{a}^\dagger\hat{a} -
    \tfrac14 \hat{a}^{\dagger 2} +\tfrac14 \hat{a}^2.
\end{split}
\end{equation}
The truncation of expansion \eqref{eq9b} is justified by assuming that
the quantum contributions $\hat{\psi}$, $\hat{\theta}$, and
$\hat{\eta}$ are ``small''. As will be shown in Sec. \ref{sec.GPE}, this
assumption permits the neglect of all terms of third order and higher
in these fields. While the assumption of small quantum fluctuations
holds in three dimensions, it fails in lower dimensions, where terms
involving products of these nominally small quantities may actually
diverge. We will address this issue in due course; for now, we naively
proceed with the expansion.

In terms of the quantities $\varphi$, $\hat{a}$ and $\hat A$, the
different terms appearing in \eqref{eq7} read
\begin{equation}\label{eq13a}
\Phi =\varphi(1+\langle \hat{A}\rangle),
\end{equation}
\begin{equation}\label{eq13b}
\tilde{n}=|\varphi|^2 \langle \hat{a}^\dagger \hat{a}\rangle, \quad
 \tilde{m}=\varphi^2 \langle \hat{a} \hat{a}\rangle,
\end{equation}
and
\begin{equation}\label{eq14}
\begin{split}
  & |\Phi|^2\Phi =|\varphi|^2 \varphi
  (1+2 \langle \hat{A}\rangle+ \langle \hat{A}^\dagger\rangle),\\
& \tilde{n} \Phi=
|\varphi|^2 \varphi \langle \hat{a}^\dagger \hat{a}\rangle,\quad
\tilde{m}\Phi^*=|\varphi|^2 \varphi \langle \hat{a} \hat{a}\rangle,
\end{split}
\end{equation}
where the quantum fields contribution has been kept only up to second order. 

Eq. \eqref{eq13a} shows that the classical fields
$\Phi(\boldsymbol{x},t)$ and $\varphi(\boldsymbol{x},t)$ are not
identical, indicating that the separation between quantum and
classical contributions in Eq. \eqref{eq2} does not exactly match the
one in Eq. \eqref{eq9a}.  The interest in the amplitude-phase
formalism lies in the fact that, whereas in low dimension $\Phi$ is an
ill-defined quantity which obeys a singular equation, we will see in
Sec. \ref{sec.BRClass} that the quantity $\varphi$ is well behaved.
Accordingly, Eq. \eqref{eq7} which governs the dynamics of the
classical field is written as
\begin{equation}\label{eq15}
\begin{split}
i\hbar\partial_t (\varphi+\varphi\langle\hat{A}\rangle)=& 
\left[ -\frac{\hbar^2}{2 m}\boldsymbol{\nabla}^2 + U-\mu\right]
(\varphi+\varphi\langle\hat{A}\rangle) \\
& +g |\varphi|^2\varphi
\left(1+2\langle\hat{A}\rangle + \langle\hat{A}^\dagger\rangle\right) \\
& + g |\varphi|^2\varphi \left(2\langle\hat{a}^\dagger\hat{a}\rangle
+ \langle\hat{a}\hat{a}\rangle\right),
\end{split}
\end{equation}
where the term $\langle\hat\psi^\dagger\hat\psi\hat\psi\rangle$ has
been dropped in accordance with our accounting for the contribution of
quantum fields only up to second order.  A technical remark is in
order here: it is possible to include this term, to keep all the terms
in expansion \eqref{eq9b} without stopping at second order, and
likewise to keep all the contributions to $\hat A$ (without
restricting oneself to the terms included in \eqref{eq12}) and then to
start the perturbative treatment only in the next section. This choice
of presentation would be indeed more logical, but it leads to
unnecessarily awkward expressions which would be immediately expanded
to second order in the next section.

\section{Perturbative expansion}\label{sec.perturb}

In this section we present the different steps of the perturbative expansion
of Eqs. \eqref{eq15} and \eqref{eq8}
leading to the backreaction equations
\eqref{BR8} and \eqref{BR9}.

\subsection{Classical fields and Gross-Pitaevskii equation}\label{sec.GPE}

At lowest order all the contributions of the quantum fields are
discarded from Eq. \eqref{eq15}, or equivalently from Eq. \eqref{eq7}. 
The corresponding approximation of the classical field $\varphi$ is the
Gross-Pitaevskii order parameter $\varphi_{\rm\sss GP}(\boldsymbol{x},t)$
solution of
\begin{equation}\label{eq16}
    i\hbar\partial_t \varphi_{\rm\sss GP}= 
    \left[ -\frac{\hbar^2}{2 m}\boldsymbol{\nabla}^2
      + U-\mu+g |\varphi_{\rm\sss GP}|^2\right] \varphi_{\rm\sss GP}.
\end{equation}
At this order, the classical fields, $\varphi$ defined in
Eq. \eqref{eq10} and $\Phi$ defined in Eq. \eqref{eq2}, are both equal
to the solution $\varphi_{\rm\sss GP}$ of the Gross-Pitaevskii equation
\eqref{eq16}.
Their modifications induced by the quantum
fluctuations appear at higher order. To take them into account we write
\begin{equation}\label{eq.expand}
\varphi=\varphi_{\rm\sss GP}+
\delta\varphi, \quad\mbox{and}\quad
\Phi=\varphi_{\rm\sss GP}+
\delta\Phi.
\end{equation}
Comparing Eqs. \eqref{eq15} and \eqref{eq16} makes it clear that the
first effects of quantum fluctuations on
$\delta\varphi(\boldsymbol{x},t)$ and $\delta \Phi(\boldsymbol{x},t)$
are of second order in $\hat{a}$ and $\hat{a}^\dagger$, which
legitimates our discarding of contributions involving terms of third
order in these fields in Sec. \ref{sec.AP}, since they arise at the
next order in perturbation.  We may thus replace Eq. \eqref{eq13a} by
\begin{equation}\label{eq17}
    \delta\Phi=\delta\varphi+\varphi_{\rm\sss GP}\langle \hat{A}\rangle.
\end{equation}
The correction to $\varphi_{\rm\sss GP}$ is accordingly computed from
an expansion of Eq. \eqref{eq15} in which are kept only the terms
linear in $\delta\varphi$ (or $\delta\Phi$) and quadratic in the
quantum fields $\hat{a}$ and $\hat{a}^\dagger$ (or equivalently
$\hat\eta$ and $\hat\theta$):
\begin{equation}\label{eq18}
(i\hbar\partial_t-{L})\delta\Phi-
g\varphi_{\rm\sss GP}^2\delta\Phi^*
= g |\varphi_{\rm\sss GP}|^2\varphi_{\rm\sss GP}
(2 \langle \hat{a}^\dagger\hat{a}\rangle + 
\langle \hat{a}\hat{a}\rangle),
\end{equation}
where
\begin{equation}\label{eq19}
  {L}=-\frac{\hbar^2}{2 m}\boldsymbol{\nabla}^2 + U-\mu
  +2 g |\varphi_{\rm\sss GP}|^2.
\end{equation}
Equation \eqref{eq18} describes the modification of the classical fields beyond the Gross-Pitaevskii approximation. It
involves terms of second order in the quantum
fields, the evaluation of which necessitates the solution of the first
order Bogoliubov equation, see Sec. \ref{sec.Pert_Bogo} below.  Some
of these terms are ill defined in low dimension. This issue will be
addressed in Sec. \ref{sec.BRClass}.

\subsection{First order: Bogoliubov equation for the quantum fields}\label{sec.Pert_Bogo}
According to our perturbative scheme, in Eqs. \eqref{eq17} and
\eqref{eq18} the quantum averages should be evaluated from the
solution of \eqref{eq8} where the classical fields are the solutions
of the Gross-Pitaesvkii equation \eqref{eq16} and only linear terms in
the quantum fields are included. This leads to the celebrated Bogoliubov
equation:
\begin{equation}\label{eq20}
(i\hbar\partial_t-{L})\hat{\psi}- g \varphi_{\rm\sss GP}^2 \hat{\psi}^\dagger=0.
\end{equation}
It is convenient for our purpose to express this equation in terms of
the amplitude and phase fields $\hat\eta$ and $\hat\theta$ defined in
Eq. \eqref{eq9a}. To this end we write the solution $\varphi_{\rm\sss
  GP}$ of the Gross-Pitaevskii equation \eqref{eq16} as
\begin{equation}\label{eq21}
    \varphi_{\rm\sss GP}(\boldsymbol{x},t)=
    \sqrt{\rho_{\rm\sss GP}(\boldsymbol{x},t)}
    \exp\{i\Theta_{\rm\sss GP}(\boldsymbol{x},t)\},
\end{equation}
so that the Gross-Pitaevskii density and velocity read
\begin{equation}\label{eq22}
  \rho_{\rm\sss GP}(\boldsymbol{x},t)=|\varphi_{\rm\sss GP}|^2,
  \quad\mbox{and}\quad
    \boldsymbol{V}_{\!\!\rm\sss GP}(\boldsymbol{x},t)=
    \frac{\hbar}{m}\boldsymbol{\nabla}\Theta_{\rm\sss GP}.
\end{equation}
From the definitions \eqref{eq2} and \eqref{eq11} we have:
\begin{equation}\label{eq23}
  \hat\psi=\hat\Psi-\langle\hat\Psi\rangle=\varphi\, \hat{a}
  +\varphi \left(\hat{A}-\langle\hat{A}\rangle\right).
\end{equation}
At the order of accuracy at which Eq. \eqref{eq20} is solved, one may
replace \eqref{eq23} by
\begin{equation}\label{eq24}
    \hat\psi\simeq \varphi_{\rm\sss GP}\,\hat{a}=\varphi_{\rm\sss GP} 
    \left( \tfrac12 \hat{\eta} + i \hat\theta\right).
\end{equation}
Then, since $\varphi_{\rm\sss GP}$ is a solution of \eqref{eq16},
$(i\hbar\partial_t-{L})\varphi_{\rm\sss GP}= -g |\varphi_{\rm\sss
  GP}|^2\varphi_{\rm\sss GP}$ and the first term in Eq. \eqref{eq20}
can be cast under the form
\begin{equation}\label{eq25}
  (i\hbar\partial_t-{L})\hat\psi=\varphi_{\rm\sss GP}
  (-g |\varphi_{\rm\sss GP}|^2+i\hbar\partial_t-{\cal L})\hat{a},
\end{equation}
where
\begin{equation}\label{eq26}
{\cal L}=
-\frac{\hbar^2}{m}
\left(\frac{\boldsymbol{\nabla}\varphi_{\rm\sss GP}}{\varphi_{\rm\sss GP}}\right)
    \cdot\boldsymbol{\nabla}-\frac{\hbar^2}{2 m}\boldsymbol{\nabla}^2\; .
\end{equation}
It follows from Eq. \eqref{eq21} that
\begin{equation}\label{eq27}
  \frac{\boldsymbol{\nabla}\varphi_{\rm\sss GP}}{\varphi_{\rm\sss GP}}=
  \tfrac12 \frac{\boldsymbol{\nabla}\rho_{\rm\sss GP}}{\rho_{\rm\sss GP}}
  + i \boldsymbol{\nabla}\Theta_{\rm\sss GP},
\end{equation}
which makes it possible to recast the operator ${\cal L}$ under the form:
\begin{equation}\label{eq28}
{\cal L}=-\frac{\hbar^2}{2 m \rho_{\rm\sss GP}}
\boldsymbol{\nabla}\cdot\rho_{\rm\sss GP}\boldsymbol{\nabla} 
-i \hbar \boldsymbol{V}_{\!\!\sss GP}\cdot \boldsymbol{\nabla} \; ,
\end{equation}
and Eq. \eqref{eq20} as
\begin{equation}\label{eq29}
  (i\hbar\partial_t-\hat{\cal L})\hat{a}=
  g|\varphi_{\rm\sss GP}|^2(\hat{a}+\hat{a}^\dagger)=g\rho_{\rm\sss GP}\hat{\eta}.
\end{equation}
Adding and subtracting Eq. \eqref{eq29} with its Hermitian 
conjugate yields 
\begin{equation}\label{eq30}
  \hbar(\partial_t+\boldsymbol{V}_{\!\!\sss GP}\cdot\boldsymbol{\nabla})
  \hat\theta -
  \frac{\hbar^2}{4 m \rho_{\rm\sss GP}}
  \boldsymbol{\nabla}\cdot(\rho_{\rm\sss GP}\boldsymbol{\nabla}\hat\eta)
  +g \rho_{\rm\sss GP}\hat\eta=0,
\end{equation}
and
\begin{equation}\label{eq31}
(\partial_t+ \boldsymbol{V}_{\!\!\sss GP}\cdot\boldsymbol{\nabla})\hat\eta
+\frac{\hbar}{m \rho_{\rm\sss GP}}
\boldsymbol{\nabla}\cdot(\rho_{\rm\sss GP}\boldsymbol{\nabla}\hat\theta)=0.
\end{equation}
Equations \eqref{eq30} and \eqref{eq31} are the amplitude-phase
versions of \eqref{eq20}. They are typically solved in cases where
$\varphi_{\rm\sss GP}$ is time-independent. We here consider the more
general situation where the density $\rho_{\rm\sss GP}$ and velocity
$\boldsymbol{V}_{\!\!\sss GP}$ may depend not only on space but also
on time.

It will appear convenient in future computations to introduce the
following operators (which act on $(\boldsymbol{x},t)$-dependent
scalar quantities)
\begin{subequations}\label{eq.ope}
\begin{align}
& \mathcal{T}= \partial_t + \boldsymbol{V}_{\!\!\sss GP}\cdot\boldsymbol{\nabla}
\label{eq.opea}
\\   
  & \mathcal{X}=\frac{\hbar}{2 m \rho_{\rm\sss GP}}
    \boldsymbol{\nabla}\cdot\rho_{\rm\sss GP}\boldsymbol{\nabla}
\label{eq.opeb}
\end{align}
\end{subequations}
In terms of operators $\mathcal{T}$ and $\mathcal{X}$ one may write
\begin{equation}\label{eq.recast}
i\hbar\partial_t -{\cal L}=\hbar \left( i \mathcal{T} + \mathcal{X}\right),
\end{equation}
and the Bogoliubov equations \eqref{eq30} and \eqref{eq31} read
\begin{subequations}\label{eq.bogo}
\begin{align}
& \tfrac12 {\cal X}\hat{\eta} - {\cal T}\hat\theta 
= \frac{g\rho_{\rm\sss GP}}{\hbar} \hat{\eta},
\label{eq.bogoa}
\\
& \tfrac12 {\cal T}\hat{\eta} + {\cal X}\hat\theta=0 .
\label{eq.bogob}
\end{align}
\end{subequations}

\subsection{Averages of quantum fields}\label{sec.averages}
In this section we give explicit expressions the quadratic averages
of the quantum fields which appear in Eq. \eqref{eq18}.

The first quantity to be considered is the density-density correlation
function $G^{(2)}$ defined as
\begin{equation}\label{eq32}
\begin{split}
  G^{(2)}(\boldsymbol{x},\boldsymbol{y},t)  = &
  \langle :\! \hat{\rho}(\boldsymbol{x},t)
  \hat{\rho}(\boldsymbol{y},t)\! : \rangle
    - \langle \hat{\rho}(\boldsymbol{x},t)\rangle  
    \langle \hat{\rho}(\boldsymbol{y},t) \rangle,\\
    = & \langle \hat{\rho}(\boldsymbol{x},t)
    \hat{\rho}(\boldsymbol{y},t) \rangle
    -\delta(\boldsymbol{x}-\boldsymbol{y}\,) \langle 
    \hat{\rho}(\boldsymbol{y},t) \rangle\\
    & - \langle \hat{\rho}(\boldsymbol{x},t)\rangle  
    \langle \hat{\rho}(\boldsymbol{y},t) \rangle
\end{split}
\end{equation}
where the operators between colons are normal ordered and $\hat\rho$
is the density operator whose exact expression (valid at all orders)
is
\begin{equation}\label{eq33}
    \hat{\rho}=\hat{\Psi}^\dagger\hat\Psi=\rho\, (1+ \hat\eta).
\end{equation}
At the order of the expansion considered in Sec. \ref{sec.Pert_Bogo}
(which is the relevant one) this yields directly
\begin{equation}\label{eq34}
  g^{(2)}(\boldsymbol{x},t)\equiv
  \frac{G^{(2)}(\boldsymbol{x},\boldsymbol{x},t)}{{\rho}^2(\boldsymbol{x},t)}
  =\langle\hat\eta^2(\boldsymbol{x},t)\rangle -
  \frac{\delta(\boldsymbol{0})}{\rho_{\rm\sss GP}(\boldsymbol{x},t)}.
\end{equation}
This expression is regular in 1D: the $\delta$ peak contribution
exactly cancels an ultraviolet divergence in
$\langle\hat\eta^2\rangle$ computed from the solution of the
Bogoliubov equation. It is ultraviolet diverging in 2D and 3D (see
below).

We will also need to evaluate the average of the combination
\begin{equation}\label{eq35}
    \hat{a}^\dagger\hat{a}+\hat{a}\hat{a}=\tfrac12 \hat\eta^2 +\tfrac{i}{2}
    \left[\hat{\eta},\hat\theta\right] +\tfrac{i}{2}
    \left(\hat{\eta}\hat\theta+\hat\theta\hat\eta\right),
\end{equation}
where the explicit $(\boldsymbol{x},t)$ dependence of all the terms
has been omitted for legibility and the second member has been
obtained by use of Eq. \eqref{eq12} and of the relation
\begin{equation}\label{eq38}
\hat\theta\hat\eta=-\tfrac12 \left[\hat\eta,\hat\theta\right] 
+\tfrac12 
\left(
\hat\eta\hat\theta+\hat\theta\hat\eta
\right).
\end{equation}
Relation \eqref{eq24} and the fact that operators $\hat\psi$ and
$\hat\psi^\dagger$ obey the standard Bose commutation relations
\eqref{eq.commu} impose\footnote{Note that the commutation relation
  \eqref{eq36} is not exact. It may be shown that it holds up to third
  order in the quantum fields, which is more than sufficient for our
  perturbative scheme.}
\begin{equation}\label{eq36}
\left[\hat{\eta}(\boldsymbol{x},t),\hat\theta(\boldsymbol{y},t)\right]
=\frac{i}{\rho_{\rm\sss GP}(\boldsymbol{x},t)} \,
\delta(\boldsymbol{x}-\boldsymbol{y}),
\end{equation}
and from \eqref{eq34} the average of \eqref{eq35} thus reads
\begin{equation}\label{eq37}
    \langle \hat{a}^\dagger\hat{a} + \hat{a}\hat{a}\rangle = 
    \tfrac12 g ^{(2)} + 
    i \Re\langle \hat\eta\,\hat\theta\rangle .
\end{equation}
This relation makes it possible to derive an expression for
$g^{(2)}(\boldsymbol{x},t)$ alternative to \eqref{eq34}:
\begin{equation}\label{eq34bis}
    g^{(2)}=\langle \hat{a}^\dagger\hat{a} + \hat{a}\hat{a}\rangle + \rm{c.c.}\; ,
\end{equation}
where ``c.c.'' stands for ``complex conjugate''.

We also need to evaluate the average of operator $\hat A$ defined in
Eq. \eqref{eq12}. Use of relation \eqref{eq38} leads to
\begin{equation}\label{eq39}
\begin{split}
\langle \hat{A}\rangle = & 
- \tfrac12 \langle \hat{a}^\dagger\hat{a}\rangle +
\tfrac14 \langle \hat{a}^2 -\hat{a}^{\dagger\,2}\rangle
\\
= & \left\langle
-\tfrac12 \hat{\theta}^2 -\tfrac18 \hat{\eta}^2
-\tfrac{i}{4}[\hat{\eta},\hat{\theta}]\right\rangle
+\tfrac{i}{4}
\left\langle
\hat\eta\hat\theta+\hat\theta\hat\eta
\right\rangle
\\
= & \left\langle
 -\tfrac12 \hat\theta^2
-\tfrac18 \hat\eta^2+
\tfrac14 \frac{\delta(\boldsymbol{0})}{\rho_{\rm\sss GP}}\right\rangle
+\tfrac{i}{2} \Re\langle \hat\eta\hat\theta\rangle\\
= & -\tfrac12 \left\langle
\hat\theta^2  -\tfrac14 \frac{\delta(\boldsymbol{0})}{\rho_{\rm\sss GP}}
\right\rangle
-\tfrac18 g^{(2)} +\tfrac{i}{2} \Re\langle \hat\eta\hat\theta\rangle,
\end{split}
\end{equation}
where Eq. \eqref{eq34} has been employed to obtain the last
expression.  The term between brackets in the third line of
\eqref{eq39} is identical to the first average in the right hand side
of the first line. It is known as the depletion of the condensate. It
is ultraviolet convergent in dimensions 1, 2 and 3, but phase
fluctuations make it infrared diverging in one dimension at zero
temperature. It is this effect which forbids {\it bona fide} Bose-Einstein 
condensation in 1D. However, it has been shown in \cite{Anderson2015} that 
this divergence is
position-independent and thus killed by the spatial derivative acting
on $\langle \hat{A}\rangle $ in Eq. \eqref{BR8} below.

Explicit expressions of quantities such as those appearing in the last
line of Eq. \eqref{eq39} are given in Appendix \ref{app.C} in terms of
a Bogoliubov expansion in the presence of a sonic horizon.

\subsection{Second order: backreaction for the classical fields}\label{sec.BRClass}

Once the Bogoliubov equation solved, we are able, thanks to the
expressions derived in previous section, to explicitly compute the
averages of quantum fields appearing in the backreaction equation
\eqref{eq18}. It is convenient here to write
$\delta\Phi=\varphi_{\rm\sss GP} \delta\Phi/\varphi_{\rm\sss GP}$ and
to use a formula exactly analogous to Eq. \eqref{eq25}:
\begin{equation}\label{BR1}
    (i\hbar\partial_t-{L})\delta\Phi=
    \varphi_{\rm\sss GP}(-g |\varphi_{\rm\sss GP}|^2+i\hbar\partial_t-
    {\cal L})\frac{\delta\Phi}{\varphi_{\rm\sss GP}},
\end{equation}
where $L$ and ${\cal L}$ are defined by Eqs. \eqref{eq19}
and \eqref{eq28}, respectively. Then, the use of
Eq. \eqref{eq17} and simple manipulations make it possible to
eliminate $\delta\Phi$ from \eqref{eq18} and to recast this equation under
the form:
\begin{equation}\label{BR2}
\begin{split}
& (i\hbar\partial_t-{\cal L})
\left(\frac{\delta\varphi}{\varphi_{\rm\sss GP}}+\langle \hat{A}\rangle\right) 
- g \rho_{\rm\sss GP} \left(
  \frac{\delta\varphi}{\varphi_{\rm\sss GP}}
  + \frac{\delta\varphi^*}{\varphi_{\rm\sss GP}^*}
\right)
\\
& = g \rho_{\rm\sss GP}\left(
2 \langle \hat{a}^\dagger\hat{a}\rangle + 
\langle \hat{a}\hat{a}\rangle
+\langle \hat{A}\rangle+\langle \hat{A}^\dagger\rangle
\right)\\
& = g \rho_{\rm\sss GP}
\langle \hat{a}^\dagger\hat{a} + 
\hat{a}\hat{a}\rangle.
\end{split}
\end{equation}
In the above expression the second line has been simplified
--resulting in the final line-- by observing that
$\langle \hat{A}\rangle+\langle \hat{A}^\dagger\rangle=-\langle
\hat{a}^\dagger\hat{a}\rangle $, as can be directly checked from
expressions \eqref{eq12}. This is an important step because in 1D the
averages $\langle \hat{a}^\dagger\hat{a}\rangle$ and
$\langle\hat{A}\rangle$ which appear in the second line of
Eq. \eqref{BR2} are ill defined due to infrared divergent phase
fluctuations, whereas the term
$\langle \hat{a}^\dagger\hat{a} + \hat{a}\hat{a}\rangle$ is infrared
regular. In 2D and 3D this source term is correctly taken into account
by a proper renormalisation of the coupling constant which cancels an
ultraviolet divergence in the anomalous average
$\langle \hat{a}\hat{a}\rangle$, see Sec. \ref{sec.Stat.BR} below or
also, e.g., Refs. \cite{Giorgini1998,Mora2003}.  Hence, whereas
Eq. \eqref{eq18} contains troublesome divergent terms, the only
possible divergence in \eqref{BR2} comes from the
$\langle\hat{A}\rangle$ term in the left hand side. However, it is
acted upon by space and time derivatives which kill its diverging part
as shown in \cite{Anderson2015} and verified below.

Note that the external potential $U(\boldsymbol{x})$ does not appear
in Eq. \eqref{BR2}. It is nonetheless implicitly included through
$\varphi_{\rm\sss GP}(\boldsymbol{x},t)$. Indeed, the fact that
$\varphi_{\rm\sss GP}$ is a solution of the Gross-Pitaevskii equation
in the presence of $U$ has been used in Eq. \eqref{BR1} for removing
the external potential from the equations. The same remark holds for
Eqs. \eqref{eq30}, \eqref{eq31} and \eqref{eq.bogo}.

In the same way as we have written
$\varphi=\varphi_{\rm\sss GP}+\delta\varphi$ in Eq. \eqref{eq.expand},
we may expand the density $\rho$ and the phase $\Theta$ defined in
\eqref{eq10} as
\begin{equation}\label{BR3}
  \rho=\rho_{\rm\sss GP} + \delta\rho,
  \quad \Theta=\Theta_{\rm\sss GP} + \delta\Theta,
\end{equation}    
where $\rho_{\rm\sss GP}$ and $\Theta_{\rm\sss GP}$ are defined in
\eqref{eq21}. This leads to
    \begin{equation}\label{BR4}
 \frac{\delta\varphi}{\varphi_{\rm\sss GP}}=\tfrac12
 \frac{\delta\rho}{\rho_{\rm\sss GP}} + i \delta\Theta.  
\end{equation}
The quantities $\delta\rho$ and $\delta\Theta$ are the modifications
of the density and the phase of the classical field induced by the
quantum fluctuations. In order to write explicitly the equations they
obey, it suffices to write the imaginary and real parts of
\eqref{BR2}.

For properly following the different steps of the computation, we 
use Eqs. \eqref{eq37} and \eqref{BR4} for
rewriting \eqref{BR2} as:
\begin{equation}\label{BR6}
\begin{split}
& (i\hbar\partial_t-{\cal L})
\left(  \tfrac12
 \frac{\delta\rho}{\rho_{\rm\sss GP}} + i \delta\Theta  
 +\langle \hat{A}\rangle\right) 
- g \delta \rho
\\
& = \tfrac12 g \rho_{\rm\sss GP}\left(
g ^{(2)} + 
    i \langle \hat\eta\,\hat\theta + \hat\theta\,\hat\eta\rangle \right) .
\end{split}
\end{equation}
Let us first consider the real part of \eqref{BR6}. From
Eq. \eqref{eq.recast} it reads
\begin{equation}\label{BR7}
\begin{split}
{\cal T}\delta\Theta 
-{\cal X} \frac{\delta\rho}{2\rho_{\rm\sss GP}}
+\frac{g\delta\rho}{\hbar} & =
{\cal X}\Re\langle\hat{A}\rangle
-{\cal T}\Im \langle\hat{A}\rangle\\
& -\frac{g \rho_{\rm\sss GP}}{2\hbar} g^{(2)}.
\end{split}
\end{equation}
Using the explicit expressions \eqref{eq.ope} and \eqref{eq39} this reads
\begin{equation}\label{BR8}
\begin{split}
  & \hbar(\partial_t+
  \boldsymbol{V}_{\!\!\sss GP}\cdot\boldsymbol{\nabla})
  \left( \delta\Theta -
  \tfrac12  \Re\langle\hat\eta\hat\theta\rangle\right)
+ g (\delta\rho +\tfrac12 \rho_{\rm\sss GP} g^{(2)}) 
\\ &
=
\\
& 
\frac{- \hbar^2}{4 m \rho_{\rm\sss GP}} \boldsymbol{\nabla}\cdot
\left[
\rho_{\rm\sss GP}\boldsymbol{\nabla} \left(
\frac{\delta\rho}{\rho_{\rm\sss GP}}
-\langle \hat{\theta}^2\rangle
+\frac{\delta(\boldsymbol{0})}{4 \rho_{\rm\sss GP}}
-\frac{g^{(2)}}{4}
\right)
\right].
\end{split}
\end{equation}
Equation \eqref{BR8} is an unsteady Bernoulli equation.  Its classical
version is usually obtained by performing a Madelung transform on the
Gross-Pitaevskii equation. We consider here the next order which
involves quadratic quantum contributions and the corresponding
modifications of the density and phase of the classical field.  In the
first line it involves the classical modification of the phase (the
velocity potential) together with a quantum fluctuation term
($\Re\langle\hat\eta\hat\theta\rangle$) and the interacting term
resulting from the non-linear dependence of pressure on density, also
affected by quantum fluctuations (the $g^{(2)}$ factor). The second
line is the contribution of the so-called quantum pressure term which
involves the modification $\delta\rho$ of the classical density and a
quantum correction corresponding to the ``quantum depletion'' $\langle
\hat{a}^\dagger\hat{a}\rangle=\langle\hat{\theta}^2\rangle - \tfrac14
\delta(\boldsymbol{0})/\rho_{\rm\sss GP}+\tfrac14 g^{(2)}$, see
Eq. \eqref{eq39}. At this point it is important to stress that the
infrared divergence of $\langle \hat{a}^\dagger\hat{a} \rangle$ in one
dimension is killed by the spatial derivative in \eqref{BR8}, a result
which has been proven in Ref. \cite{Anderson2015}. Thus
Eq. \eqref{BR8} can {\it a priori} be solved in any dimension, as
illustrated below.

It is shown in Appendix \ref{app.B} that the imaginary part of
\eqref{BR6} can be cast in the form of a continuity equation
\begin{equation}\label{BR9}
\partial_t \delta\rho +
\boldsymbol{\nabla}\cdot\left(
\boldsymbol{V}_{\!\!\sss\rm GP} \, \delta \rho + \rho_{\rm\sss GP} \delta\boldsymbol{V}
+\Re\langle \rho_{\rm\sss GP} \hat{\eta} \hat{\boldsymbol{v}}\rangle\right)=0,
\end{equation}
where
\begin{equation}\label{BR10}
\delta \boldsymbol{V}=\frac{\hbar}{m}\boldsymbol{\nabla}\delta\Theta,
\quad\mbox{and}\quad
\hat{\boldsymbol{v}}=\frac{\hbar}{m}
\boldsymbol{\nabla}\hat\theta.
\end{equation}
Eq.  \eqref{BR9} is of the form 
\begin{equation}\label{eq:conserving}
\partial_t\delta\rho + \boldsymbol{\nabla}\cdot \delta\boldsymbol{J}=0,
\end{equation} 
where
\begin{equation}\label{eq.deltaJ}
\delta\boldsymbol{J}(\boldsymbol{x},t)=
\boldsymbol{V}_{\!\!\sss GP} \delta \rho + \rho_{\rm\sss GP} \delta\boldsymbol{V}
+\Re\langle \rho_{\rm\sss GP} \hat{\eta} \hat{\boldsymbol{v}}\rangle
\end{equation}
represents the modification of the classical conserved current induced
by backreaction effects. The various contributions to $\delta
\boldsymbol{J}$ illustrate that the classical current is influenced
not only by the variations $\delta \rho$ and $\delta \boldsymbol{V}$
in the classical density and velocity fields, but also by the term
$\Re\langle \rho_{\rm\sss GP} \hat{\eta} \hat{\boldsymbol{v}}\rangle$,
which arises from quantum corrections. This contribution corresponds
to the expectation value
$\langle\;\hat{\!\boldsymbol{\jmath}}\,\rangle$ of the quantum current
operator
\begin{equation}\label{eq.Qcurrent}
  \hat{\!\boldsymbol{\jmath}}(\boldsymbol{x},t)=\tfrac12
  (\hat\rho\hat{\boldsymbol{v}} + \hat{\boldsymbol {v}}\hat{\rho})
\end{equation}
evaluated at the order appropriate to our second order expansion.

Equations \eqref{BR8} and \eqref{BR9} are the main result of the
present paper. They describe the dynamics of the leading order
modification of the classical background flow induced by quantum
fluctuations. They are valid even in a time-dependent flow for which
the zeroth order Gross-Pitaevskii equation \eqref{eq16} and the first
order Bogoliubov equations \eqref{eq20} (together with their
solutions) are explicitly position and time-dependent. They constitute
an extension of the results obtained by Mora and Castin in
Ref. \cite{Mora2003} to a possibly non-stationary flow in the presence
of a background current.  The amplitude-phase formalism makes it
possible, without using a number conserving approach, to circumvent
the infrared divergence problem encountered in other studies
\cite{Baak2022,Pal2024}.  Our treatment pertains to the
  so-called time-dependent Hartree-Fock-Bogoliubov scheme (see, e.g.,
  discussions in Ref. \cite{GNZ2009}). Note however, that we do not
  push the development as far as to include perturbatively coupled
  time-dependent equations for the condensate and noncondensate
  components of the Bose gas, as done by Giorgini in
  Ref. \cite{Giorgini2000} for instance.

\section{Stationary configurations}\label{sec.Stat.BR}

In the following we will consider a situation where the zeroth order
Gross-Pitaevskii flow is stationary, described by time-independent
density $\rho_{\rm\sss GP}(\boldsymbol{x})$ and velocity
$\boldsymbol{V}_{\rm\sss GP}(\boldsymbol{x})$. We look for a
stationary solution of the backreaction equations \eqref{BR8} and
\eqref{BR9} for which the quantities $\delta \rho$ and
$\delta \boldsymbol{V}$ are also time-independent. In such a
configuration $\delta\Theta$ is nonetheless typically time-dependent and we
introduce the quantity
\begin{equation}\label{eqs1}
    \delta \mu=-\hbar \partial_t \delta\Theta .
\end{equation}
Then
Eqs. \eqref{BR8} and \eqref{BR9} simplify to:
\begin{equation}\label{eqs2}
\begin{split}
& m \boldsymbol{V}_{\!\!\sss GP}\cdot\delta\boldsymbol{V} 
- 
\frac{\hbar^2}{4 m \rho_{\rm\sss GP}}\boldsymbol{\nabla}\cdot\left(
  \rho_{\rm\sss GP}\boldsymbol{\nabla}\frac{\delta\rho}{\rho_{\rm\sss GP}}
\right)+ g \delta\rho = \\
& \delta\mu - \tfrac12 g\rho_{\rm\sss GP} g^{(2)}
-\frac{\hbar}{2}\boldsymbol{V}_{\!\!\sss GP}\cdot\boldsymbol{\nabla}
\Re\langle\hat\eta\hat\theta\rangle
\\
& -
\frac{\hbar^2}{4 m \rho_{\rm\sss GP}} \boldsymbol{\nabla}\cdot\left[
\rho_{\rm\sss GP}\boldsymbol{\nabla} \left(
\langle \hat{\theta}^2\rangle
-\tfrac14 \frac{\delta(\boldsymbol{0}\,)}{\rho_{\rm\sss GP}}
+\tfrac14 g^{(2)}
\right)
\right],
\end{split}
\end{equation}
and
\begin{equation}\label{eqs3}
\boldsymbol{\nabla}\cdot\left(
\boldsymbol{V}_{\!\!\sss\rm GP} \, \delta \rho
+ \rho_{\rm\sss GP} \delta\boldsymbol{V}
+\Re\langle \rho_{\rm\sss GP} \hat{\eta} \hat{\boldsymbol{v}}\rangle\right)=0.
\end{equation}
It is easily verified that $\delta\mu$ is position-independent for
such stationary solutions since, from the definitions \eqref{BR10} and
\eqref{eqs1}, $\boldsymbol{\nabla}\delta\mu= -m\partial_t
\delta\boldsymbol{V}$. Besides, Eq. \eqref{eqs2} proves that
$\delta\mu$ is also time-independent, since all the other
contributions in this equation are. $\delta\mu$ is thus a constant, it
is the modification of the chemical potential induced by the
backreaction equations.

\subsection{Uniform and stationary configurations}\label{sec.BR.uni.statio}

As a simple test of the validity of the backreaction equations
\eqref{eqs2} and \eqref{eqs3}, we first consider a stationary and
uniform system (with $U(\boldsymbol{x})=0$) at zero temperature in the
absence of current in dimension $d$.  In such a system
$\boldsymbol{V}_{\rm\sss
  GP}=0=\langle\;\hat{\!\boldsymbol{\jmath}}\,\rangle$ and all the
ingredients of Eq. \eqref{eqs2} (such as $\rho_{\sss\rm GP}$,
$g^{(2)}$, $\delta\rho$, ...) are time and position independent.
Equation \eqref{eq16} reads $\mu_{\rm\sss GP}=g \rho_{\rm\sss GP}$
where $\mu_{\rm\sss GP}$ is the value of the chemical potential of a
system of constant density $\rho_{\rm\sss GP}$ according to the
Gross-Pitaevskii approach.  Likewise, Eq. \eqref{eqs2} reduces to
$\delta\mu = g\delta\rho + \tfrac12 g\rho_{\rm\sss GP}
g^{(2)}$. Combining these two results yields (noting $\mu=\mu_{\rm\sss
  GP}+\delta\mu$ and $\rho=\rho_{\rm\sss GP}+\delta\rho$)
\begin{equation}\label{eq.mu1}
\mu= g \rho +   \tfrac12 g\rho_{\rm\sss GP} g^{(2)}\; . 
\end{equation}
This is known as the Hartree-Fock-Bogoliubov result for the chemical
potential \cite{Griffin1996}. This expression differs from the
Hugenholtz-Pines result \cite{Hugenholtz1959} indicating that our
approach, if pursued at higher order, would yield a gaped spectrum: in
terms of the Hohenberg-Martin classification the present approach is
conserving and not gapless, see discussions in
\cite{Hohenberg1965,Griffin1996}. However, it is perfectly legitimate
for our purpose, as illustrated in sections \ref{sec.3dsu} and
\ref{sec.1dsu} (see also Refs. \cite{Giorgini1998,Mora2003}).
Moreover, the violation of the Hugenholtz-Pines theorem is not
necessarily prohibitive. A possible approach to recover a gapless
spectrum—along with a corrected phonon velocity and damping rate—is to
analyze the linear response of the classical field within a
perturbative framework, as demonstrated by Giorgini in
Ref. \cite{Giorgini2000}.

\subsubsection{The three dimensional case}\label{sec.3dsu}

In 3 dimensions, the last term of the right hand side of expression
\eqref{eq.mu1} has an utraviolet divergence, associated to the
anomalous average $\langle \hat{a}\hat{a}\rangle$ in
\eqref{eq34bis}. But, expanding the coupling constant to second order
in the scattering length regularizes this divergence (see e.g.,
\cite{Pitaevskii2016,Pethick2011}), leading to the Lee-Huang-Yang
expression \cite{Lee1957,Beliaev1958b}
\begin{equation}\label{eq.mu2}
    \mu = g \rho + \frac{4 g}{3\pi^2\xi^3}=
    g\rho \left[ 1 + \frac{32}{3\sqrt{\pi}}\sqrt{a^3\rho}\, \right],
\end{equation}
where 
\begin{equation}\label{eq.xi}
\xi=\frac{\hbar}{\sqrt{m g \rho}}
\end{equation}
is the healing length and $a$ the 3D
scattering length with, at leading order, $g=4\pi\hbar^2 a/m$.

\subsubsection{The one dimensional case}\label{sec.1dsu}

We are mainly interested in 1D configurations where the situation is 
actually simpler because in this case $g^{(2)}$ given from the Bogoliubov
expression \eqref{eq34bis} is regular. We get \cite{Gangardt2003}
\begin{equation}\label{eq.g21d}
    g^{(2)}=-\frac{2}{\pi\rho_{\rm\sss GP} \xi_{\rm\sss GP}},
\end{equation}
leading from \eqref{eq.mu1} to
\begin{equation}\label{eq.mu3}
    \mu =g\rho - \frac{g}{\pi\xi}.
\end{equation}
In this expression we replaced in the corrective term the
Gross-Pitaevskii healing length $\xi_{\rm\sss GP}$ by $\xi$, which is
legitimate. The result \eqref{eq.mu3} has already been obtained in
Ref. \cite{Mora2003} and corresponds to the weak interaction expansion
of the exact Lieb-Liniger result \cite{Lieb1963a}. Inserted in the
thermodynamic relation $mc^2=\rho(\partial\mu/\partial\rho)$ it yields
\begin{equation}\label{eq.mu4}
    c=\frac{1}{\sqrt{m}}\sqrt{g\rho-\frac{g}{2\pi\xi}}.
\end{equation}
This formula defines the speed of sound $c$ as a function of the
density $\rho$ in a 1D homogeneous system at zero temperature. It
corrects the Gross-Pitaevskii formula $c_{\rm\sss GP}=(g \rho_{\rm\sss
  GP}/m)^{1/2}$. It is clear from \eqref{eq.mu3} that in 1D the
dimensionless small parameter of our approach is $(\rho\xi)^{-1}$,
implying that our results are expected to be valid only in the limit
$\rho\xi\gg 1$. However, it was observed by Lieb \cite{Lieb1963b} that
expression \eqref{eq.mu4} agrees very well with the exact result down
to $\rho\xi\sim 0.3$. In the case of interest for us, a typical
experimental order of magnitude is $\rho\xi \approx 30 \leftrightarrow
60$ \cite{Steinhauer2016}, i.e., quite far from the regime where
\eqref{eq.mu4} becomes incorrect.

\subsection{Analogous black hole in a 1D stationary transonic flow}\label{sec.analog.stat.1D}

We now come to the main interest of our study: backreaction effects in
a 1D flow mimicking a black hole. Different black hole configurations
have been proposed in the past and we focus in the present work on the
ones presented in Appendix \ref{app.A}, which we denote as
``waterfall'', ``$\delta$-peak'' and ``flat profile''.

Since we work in 1D, from now on we do not consider vectors, but their
unique Cartesian coordinate. For instance we no longer use
$\boldsymbol{\nabla}$ and $\delta \boldsymbol{V}$ but $\partial_x$ and
$\delta V$ instead.  Then Eqs. \eqref{eqs2} and \eqref{eqs3} read
\begin{equation}\label{eqs4}
\begin{split}
& m V_{\rm\sss GP} \delta V - \frac{\hbar^2}{4 m \rho_{\rm\sss GP}}
\partial_x
\left(\rho_{\rm\sss GP}\partial_x\frac{\delta\rho}{\rho_{\rm\sss GP}}\right)
+g \delta\rho= \\
&
\delta\mu -\tfrac12 g\rho_{\rm\sss GP} g^{(2)} 
-\frac{\hbar}{2}V_{\rm\sss GP}\partial_x
\Re\langle\hat\eta\hat\theta\rangle \\
&
-
\frac{\hbar^2}{4 m \rho_{\rm\sss GP}} \partial_x\left(
\rho_{\rm\sss GP}
\partial_x \left(
\langle \hat{\theta}^2\rangle
-\tfrac14 \frac{\delta(0)}{\rho_{\rm\sss GP}}
+\tfrac14 g^{(2)}\right)
\right),
\end{split}
\end{equation}
and
\begin{equation}\label{eqs5}
V_{\rm\sss GP}\delta\rho+\rho_{\rm\sss GP}\delta V +  
\Re\langle \rho_{\rm\sss GP} \hat{\eta} \hat{v}\rangle
=\delta J.
\end{equation}
Whereas all the quantities in \eqref{eqs4} and \eqref{eqs5} (such as
$\delta\rho(x)$, $\rho_{\rm\sss GP}(x)$, $g^{(2)}(x)$, ...) are
position-dependent fields, $\delta\mu$ and $\delta J$ are time {\it
  and} position independent.

In the configurations we consider (waterfall, $\delta$-peak or flat
profile, see Sec. \ref{app.A}), there are two asymptotic regions, the
far upstream and the far downstream one, for which $\rho_{\rm\sss
  GP}$, $V_{\rm\sss GP}$ and all the source terms in \eqref{eqs4} and
\eqref{eqs5} are not only time, but also position-independent. The
relevant quantities being:
\begin{equation}\label{eqs6}
\begin{split}
\rho_{\rm\sss GP}^\alpha & =\lim_{x\to\pm\infty} \rho_{\rm\sss GP}(x),\quad
g^{(2)}_\alpha =\lim_{x\to\pm\infty}g^{(2)}(x),\\
V_{\rm\sss GP}^\alpha& =\lim_{x\to\pm\infty} V_{\rm\sss GP}(x),\quad
j_\alpha = 
\lim_{x\to\pm\infty}
\Re\langle \rho_{\rm\sss GP}\hat{\eta} \hat{v}\rangle(x),
\end{split}
\end{equation}
where the index $\alpha=d$ ($u$) in the limit $x\to +\infty$
($-\infty$). Finally, the  asymptotic Gross-Pitaevskii speed of sound and healing length are denoted as $c_{\rm\sss GP}^\alpha=(g \rho_{\rm\sss GP}^\alpha/m)^{1/2}$ and $\xi_{\rm\sss GP}^\alpha=\hbar/m c_{\rm\sss GP}^\alpha$, respectively. 
  
In the asymptotic regions, the backreaction equations assume a quite simple form. In particular the modification of the density is governed by the following equation:
\begin{equation}\label{{eq.wd4}}
-\frac{\hbar^2}{4 m}\partial_x^2 \delta\rho +
m \left[(\cgp^\alpha)^2-(\Vgp^\alpha)^2\right] \delta \rho = S_\alpha,
\end{equation}
where
\begin{equation}\label{eq.nj2}
S_\alpha=\rhogp^\alpha
\left[\delta\mu -\tfrac12 m(\cgp^\alpha)^2 g_\alpha^{(2)}\right]
+ m \Vgp^\alpha [j_\alpha -\delta J]
\end{equation}
is a constant source term. The corresponding solutions are of the form
\begin{equation}\label{eq.nj3}
\delta\rho(x)=\begin{cases}
\delta\rho_u+{\cal A}_u \exp({\kappa_u x}) 
     & \mbox{when } x\to-\infty,\\
\delta\rho_d+{\cal A}_d\sin(\kappa_d x +\phi)
    & \mbox{when } x\to+\infty,\\
\end{cases}
\end{equation}
where
\begin{equation}\label{eq.nj4}
\delta\rho_\alpha=\frac{S_\alpha}{m \left[(\cgp^\alpha)^2-(\Vgp^\alpha)^2\right]},
\end{equation}
and
\begin{equation}\label{eq.nj5}
\kappa_\alpha=\frac{2 m}{\hbar}\, \Big|(\cgp^\alpha)^2-(\Vgp^\alpha)^2\Big|^{1/2}.
\end{equation}
Expression \eqref{eq.nj4} faces the risk of divergence when the Mach
number $\Mgp^\alpha=\Vgp^\alpha/\cgp^\alpha$ tends to 1.  However, it
holds also in the homogeneous case for which no such
velocity-dependent divergence should occur, since, by a Galilean
transform it is always possible to work in a reference frame where the
flow velocity cancels, which is the situation studied in
Sec. \ref{sec.1dsu}. This implies that, in the specific situations
considered below, the source term \eqref{eq.nj2} should also cancel
when $\Mgp^\alpha$ tends to unity.  We will make this check in due
time [cf. Eqs. \eqref{eq.asymWu} and \eqref{eq.asymWd}].

The asymptotic modifications of the velocity are of a form similar to that of 
the density
\begin{equation}\label{eq.nj6}
\delta V(x)=\begin{cases}
\delta V_u+{\cal B}_u \exp({\kappa_u x}) 
     & \mbox{when } x\to-\infty,\\
\delta V_d+{\cal B}_d\sin(\kappa_d x +\phi)
    & \mbox{when } x\to+\infty,\\
\end{cases}
\end{equation}
where ${\cal B}_\alpha=-{\cal A}_\alpha \Vgp^\alpha/\rhogp^\alpha$ and
\begin{equation}\label{eq.nj7}
\delta V_\alpha=\frac{1}{\rhogp^\alpha}
(\delta J -j_\alpha -\Vgp^\alpha \delta\rho_\alpha).
\end{equation}
The asymptotic expressions \eqref{eq.nj3} and \eqref{eq.nj6} show
that, whereas the upstream density and velocity modifications tend to
constant values, the downstream asymptotic profile typically displays
undulations with a wave vector $\kappa_d$ associated to a zero energy
channel (identified in Fig. \ref{fig:dispersion} in Appendix
\ref{app.modes}). Such stationary undulations have been predicted to
appear in the supersonic region of a white hole configuration
non-linearly stimulated by an external seed impinging from the
subsonic region \cite{Mayoral2011}. A similar stimulated process is
not possible in a black hole configuration because the group velocity
of the corresponding zero mode is directed outward the supersonic
region\footnote{In the terminology of Appendix \ref{app.modes}, the
zero energy channel is not outgoing, see Fig. \ref{fig:dispersion}.},
and thus cannot be excited by a source located outside the black
hole. In our case, although we consider a black hole configuration,
the appearance of undulations is however perfectly legitimate because
it results from a {\it spontaneous} process, for which the effective
``seed'' (the source term $S_d$ which accounts for quantum fluctuations)
exists throughout the whole supersonic region.

The precise asymptotic behavior of $\delta\rho(x)$ and $\delta V(x)$
depends on the constants ${\cal A}_\alpha$ and $\phi$ in
\eqref{eq.nj3} and \eqref{eq.nj6} which can only be determined by a
full solution of the backreaction equations \eqref{eqs4},
\eqref{eqs5}.  This, in turn, necessitates a determination of the
source terms in the whole physical space, which can be accurately done
only by correctly taking account of zero-mode solutions of the
Bogoliubov equations \cite{Mora2003}. It has been shown in
Ref. \cite{Isoard2020} that this is indeed crucial for accurately
determining the local density-density correlation function
$g^{(2)}(x)$ in the vicinity of the horizon. However, the
contributions of the zero-modes are not necessary when considering the
upstream and downstream asymptotic regions, far away from the
horizon. This noticeably simplifies the computation of the
contribution of the quantum fluctuations to the asymptotic local
density-density correlation function $g_\alpha^{(2)}$ and to the
asymptotic current $j_\alpha$.  We show in Appendix \ref{app.C} that
they are of the form
\begin{subequations}\label{eq12def}
\begin{align}
  g_\alpha^{(2)}=& \frac{1}{\rho_{\sss\rm GP}^\alpha \xi_{\sss\rm GP}^\alpha}
  \left(-\frac{2}{\pi} 
  + {\cal G}^{(H)}_\alpha\right) , \label{eq12defa}
  \\
{j_\alpha}
=& \frac{c_{\sss\rm GP}^\alpha}{\xi_{\sss\rm GP}^\alpha} \,
{\cal J}^{(H)}_\alpha.
\label{eq12defb}
\end{align}
\end{subequations}
These expressions encompass the standard quantum fluctuation
\eqref{eq.g21d} of $g_\alpha^{(2)}$ together with additional terms
denoted as ${\cal G}^{(H)}_\alpha$ and ${\cal J}^{(H)}_\alpha$ which
are induced by the quantum Hawking radiation and cancel in the absence
of sonic horizon. These terms are determined numerically as detailed
in Eqs. \eqref{eqB1}, \eqref{eqB2}, \eqref{eqB4} and \eqref{eqB5}.

In the present pilot study we focus on the mean values
$\delta\rho_\alpha$ and $\delta V_\alpha$ of the asymptotic behaviors
\eqref{eq.nj3} and \eqref{eq.nj6} and defer a full numerical solution
of the backreaction equations to a future work.  It should be kept in
mind that, because of the downstream undulations, the modification
$\delta \rho_d$ and $\delta V_d$ are average quantities.  The
situation is simpler in the upstream region where the asymptotic
profile is flat. Hence, it is more appropriate to solve
Eqs. \eqref{eq.nj4} and \eqref{eq.nj7} fixing boundary conditions in
the upstream region where the modified flow pattern is asymptotically
featureless (constant density and velocity).  In an equilibrium
situation, in the absence of background flow velocity, it was argued
in Ref. \cite{Mora2003} that the appropriate condition is
$\delta\mu=0$, which corresponds to a grand canonical situation. In an
homogeneous system, in the presence of a constant background flow,
Galilean invariance furthermore imposes that the beyond mean-field
effects do not modify the flow velocity.\footnote{Consider a uniform
flow of constant density and velocity in the laboratory frame. In the
comoving frame of the fluid, the Gross–Pitaevskii background velocity
vanishes. Since backreaction —i.e., beyond mean-field corrections—
cannot generate a spontaneous flow in this frame, it follows that in
the laboratory frame the initial velocity remains unaffected by
backreaction.}  Because of the asymptotic homogeneity of the upstream
flow we thus impose the similar condition $\delta V_u=0$.  From
Eqs. \eqref{eq.nj2}, \eqref{eq.nj4} and \eqref{eq.nj7} these two
conditions determine $\delta J$, $\delta\rho_u$, $\delta\rho_d$ and
$\delta V_d$.

\begin{figure}
\centering
\includegraphics[width=0.99\linewidth]{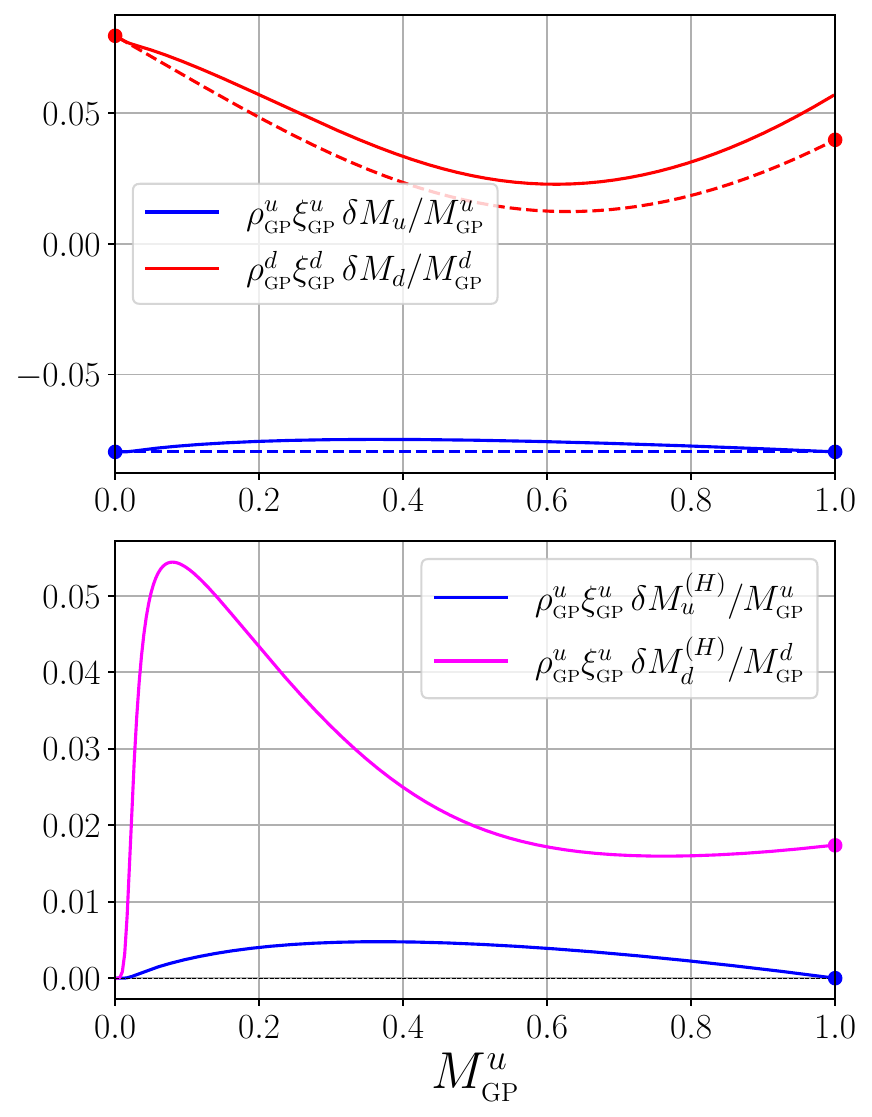}
\caption{Effect of backreaction on the asymptotic Mach numbers for the
  waterfall configuration. The results are plotted as functions of the
  upstream asymptotic Mach number (computed within the
  Gross-Pitaevskii approximation) $\Mgp^u$.\newline Upper plot: the
  solid lines represent the rescaled relative modifications $\delta
  M_u/\Mgp^u$ and $\delta M_d/\Mgp^d$ of the upstream and downstream
  Mach numbers. The dashed lines terminated by circles present the
  result obtained when discarding the contribution of the Hawking
  radiation. \newline Lower plot: relative variation of the Mach
  numbers computed by removing the usual beyond mean field
  contributions and incorporating only the backreaction induced by
  Hawking radiation. The values of the functions for $\Mgp^u=1$ are
  marked with full circles and correspond to expressions
  \eqref{eq.asymWu} and \eqref{eq.asymWd}.  }\label{fig:W}
\end{figure}

In the framework of the Gross-Pitaevskii approximation, it has been
shown that the main characteristics of the analog black hole and of
the associated Hawking radiation are determined by the values of the
asymptotic Mach numbers $\Mgp^\alpha=\Vgp^\alpha/\cgp^\alpha$ (see,
e.g., Ref. \cite{Larre2012}).  It is thus important to evaluate to
which extent these quantities are affected by the modifications
$\delta\rho_\alpha$ and $\delta V_\alpha$ of the asymptotic
densities. Backreaction effects modify the Gross-Pitaevskii result to
$M_\alpha=M^\alpha_{\rm\sss GP}+\delta M_\alpha$ with
\begin{equation}\label{eq.dMsM}
\frac{\delta M_\alpha}{M^\alpha_{\rm\sss GP}} =
\frac{\delta V_\alpha}{V^\alpha_{\rm\sss GP}} -
\frac{\delta c_\alpha}{c^\alpha_{\rm\sss GP}}
=\frac{\delta V_\alpha}{V^\alpha_{\rm\sss GP}} -\tfrac12 
\frac{\delta \rho_\alpha}{\rho^\alpha_{\rm\sss GP}}
+\frac{1}{4\pi\xi^\alpha_{\rm\sss GP}\rho^\alpha_{\rm\sss GP}},
\end{equation}
where use has been made of expression \eqref{eq.mu4} for the
one-dimensional speed of sound. In the upper panel of
Figs. \ref{fig:W} we display the relative modifications $\delta
M_\alpha/\Mgp^\alpha$ for the waterfall configuration (the results for
other configurations are presented in Appendix \ref{app.E}).  In
region $\alpha$ ($\alpha=u$ or $d$) this modification is proportional
to the small parameter $(\rhogp^\alpha\xigp^\alpha)^{-1}$, the value
of which depends of the experimental situation, and is typically of
order 0.02 in Steinhauer's experiment \cite{Steinhauer2016}.  This is
the reason why $\delta M_\alpha/\Mgp^\alpha$ is rescaled by
$\rhogp^\alpha\xigp^\alpha$ in the upper panel of this figure. In the
same plot, the result obtained when discarding the contribution of
Hawking radiation [i.e., by removing the contributions ${\cal
    G}_\alpha^{(H)}$ and ${\cal J}_\alpha^{(H)}$ in
  Eqs. \eqref{eq12def}] are represented by dashed lines.  The blue
horizontal dashed line corresponds to the value $-1/4\pi$ which would
be the rescaled relative modification of the Gross-Pitaevskii Mach
number of a homogeneous condensate (constant density $\rhogp^u$ and
constant velocity $\Vgp^u$) induced by beyond mean-field
effects. Concerning the downstream modification $\delta M_d$ of the
Mach number, we stress that it is evaluated from the average
quantities $\delta V_d$ and $\delta \rho_d$. When the period
$2\pi/\kappa_d$ of the undulations of the downstream density is small
compared to $\xigp^d$ (i.e., when $\Mgp^d\gg 1$) it is {\it a priori}
not possible to define a local speed of sound: in this limit $\delta
M_d$ defined by \eqref{eq.dMsM} is not the spatial average of an
hypothetical local downstream Mach number.

The primary observation drawn from the upper panel of Figure
\ref{fig:W} is that the relative modifications in $M_u$ and $M_d$ are
small.  The upstream Mach number weakly decreases and the downstream
one weakly increases. It thus appears that, for the chosen boundary
conditions, the transonic character of the flow is not modified by
quantum fluctuations and backreaction. The flows are poorly affected
in the whole range $0< \Mgp^u<1$, which legitimates our perturbative
approach.  Another clear feature is that the solid lines depart weakly
from the dashed ones.  This means that Hawking radiation has a lesser
impact on the Mach number than the standard beyond mean-field quantum
fluctuations. However, one should keep in mind that the beyond
mean-field corrections are present even in the absence of an acoustic
horizon, i.e., they should be taken into account already in a
homogeneous system, before the formation of the analog black hole. It
is thus appropriate to distinguish the beyond-mean field terms from
the backreaction effects truly induced by Hawking radiation.  This is
the reason why we represent in the lower panel of Fig. \ref{fig:W} the
quantities $\delta M_\alpha^{(H)}$ ($\alpha=u$ or $d$) which are the
difference between the modification of the Mach number computed with
and without the Hawking contribution. For being able to compare the
relative upstream and downstream influence of Hawking radiation, in
the lower panels we rescale $\delta M_u^{(H)}/\Mgp^u$ and $\delta
M_d^{(H)}/\Mgp^d$ by the same quantity $\rhogp^u\xigp^u$. It appears
that Hawking radiation induces positive modification of the upstream
and downstream Mach numbers, of roughly the same order. We show in
Appendix \ref{app.E} that, while the positivity of $\delta M_u^{(H)}$
observed in the lower panel of Fig. \ref{fig:W} is a general feature
due to our specific choice of boundary conditions, $\delta M_d^{(H)}$
may be positive or negative in other configurations.

It is also important to stress that in the limit $\Mgp^u\to 1$ [which
  also imposes $\Mgp^d\to 1$ from \eqref{eq.wd4}] the source term in
\eqref{eq.nj4} cancels and thus $\delta\rho_\alpha$, $\delta V_\alpha$
and $\delta M_\alpha$ do not diverge. Simple algebraic manipulations
and the use of expression \eqref{eqDL} show that
\begin{equation}\label{eq.asymWu}
    \lim_{\Mgp^u\to 1} \rhogp^u\xigp^u\frac{\delta M_u}{\Mgp^u}=-\frac{1}{4\pi},
\end{equation}
and that for the waterfall configuration
\begin{equation}\label{eq.asymWd}
    \lim_{\Mgp^u\to 1} \rhogp^d\xigp^d\frac{\delta M_d}{\Mgp^d}=\frac{1}{8\pi}
    + \frac{9}{8}\left( {\cal H}_u -{\cal H}_d\right),
\end{equation}
where ${\cal H}_u$ and ${\cal H}_d$ are defined in \eqref{eqDL}.  This
regular behavior can be attributed to two factors: the cancellation of
the source term \eqref{eq.nj2} in the limit of unit Mach number [with
  the Hawking contribution likewise canceling, as noted in
  \eqref{eqDL}], and the fact that $\Mgp^d$ approaches 1 as $\Mgp^u$
does.

\section{Conclusion and perspectives}\label{sec.conclusion}

In this work we have obtained backreaction equations [Eqs. \eqref{BR8}
  and \eqref{BR9}] describing the effect of quantum fluctuations onto
the background flow in a Bose-Einstein condensate.  These equations
have been derived in arbitrary dimension $d$, for a possibly
current-carrying, or non-stationary background, in the limit where the
quantity $\rho\xi^d$ is large compare to unity.

Our main interest lies in 1D situations with a transonic flow
realizing a sonic horizon. In this case, we solved the stationary
backreaction equations far from the acoustic horizon (deep in the
upstream and downstream regions). Whereas the asymptotic upstream flow
tends to a homogeneous limit, the existence of downstream channels of
zero energy (themselves resulting from the transonic character of the
flow) induces, through backreaction effects, stationary density and
velocity undulations in the downstream region. The existence of
undulations in the interior of the analog black hole is a nonlinear
effect of quantum backreaction, not expected within a Gross-Pitaevskii
approach.

A natural extension of our work is to consider a generic situation,
such as the ones experimentally realized in
Refs. \cite{Steinhauer2016,deNova2019}, and to determine a stationary
solution of the backreaction equations, not only in the asymptotic
regions, but in the whole physical space. In this case, the
theoretical expansion of the quantum fluctuations should explicitly
account for the existence of zero-modes of Bogoliubov equations
\cite{Blaizot1986}.  In such generic situations, on the basis of the
orders of magnitudes we obtained in the asymptotic regions, we expect
that the backreaction effects should be small, and would not
drastically affect the leading order Gross-Pitaevskii flow profile.

It should be noted that the detailed form of the stationary solutions
and of the modifications of the upstream and downstream Mach numbers
depend on the boundary conditions imposed at infinity. Simple physical
arguments lead to impose $\delta\mu=0=\delta V_u$, but other more
elaborate conditions may be imposed which modify the specifics of the
results presented in Figs. \ref{fig:W}, \ref{fig:DP}, \ref{fig:FP_W}
and \ref{fig:FP_DP} (such as the positivity of $\delta M_u^{(H)}$ for
instance).  Since the backreaction equations we derived are valid for
a time-dependent background flow, a natural way to circumvent this
issue would be to study the dynamics of formation of an acoustic
horizon in a quasi 1D BEC, as already studied in
Refs. \cite{Carusotto2008,Kamchatnov2012,Michel2015,Wang2017,Tettamanti2020,Kolobov2021,Fabbri2021},
with account of quantum backreaction effects.  Work in this direction
is in progress.

\begin{acknowledgments}
We are very grateful to M. Isoard for fruitful discussions at crucial
stages of this work. N. P. acknowledges insightful exchanges with
S. Giorgini.  A. F. acknowledges partial financial support by the
Spanish Grant PID2023-149560NB-C21 funded by
MCIN/AEI/10.13039/501100011033 and FEDER, European Union, and the
Severo Ochoa Excellence Grant CEX2023-001292-S. G. C. acknowledges
partial financial support by the Deutsche Forschungsgemeinschaft
funded Research Training Group “Dynamics of Controlled Atomic and
Molecular Systems” (Grant No. RTG 2717).
\end{acknowledgments}

\begin{appendix}
\section{Different black-hole configurations}\label{app.A}

In this study, we examine three different types of analog black holes,
referred to as ``waterall'', ``$\delta$-peak'' and ``flat profile'' in
Ref. \cite{Larre2012}.  They all correspond to the flow of a
one-dimensional BEC which is subsonic in the upstream region and
supersonic downstream.  The associated order parameter is a stationary
solution of the Gross-Pitaevskii equation \eqref{eq16}. According to
the convention adopted in this work, the corresponding fields and
characteristic quantities should be denoted with a subscript
``GP''. However, to facilitate readability, we omit this index in this
appendix.

In these configurations the background classical field is of the form:
\begin{equation}\label{eq.wd1}
    \Phi(x)=\sqrt{\rho_\alpha} \, \phi_\alpha(x) \exp(i k_\alpha x),
\end{equation}
where the index $\alpha$ takes the value $\alpha=u$ for $x<0$
(upstream region) and $\alpha=d$ for $x>0$ (downstream region).
$k_\alpha=m V_\alpha/\hbar$ where $V_\alpha>0$ is the asymptotic
velocity of the flow (in the limit $x\to-\infty$ if $\alpha=u$ and for
$x>0$ if $\alpha=d$).  We have $\lim_{x\to\pm\infty}|\phi_\alpha|=1$
and $\rho_\alpha$ is thus the asymptotic density in region
$\alpha$. The asymptotic speed of sound, Mach number and healing
length are $c_\alpha=(g\rho_\alpha/m)^{1/2}$,
$M_\alpha=V_\alpha/c_\alpha$ and $\xi_\alpha=\hbar/m c_\alpha$,
respectively.

Let us first present the specifics of the $\delta$-peak and waterfall
configurations.  In both configurations the chemical potential is
$\mu=\tfrac12 m V_u^2+g\rho_u$ and upstream and downstream quantities
are related through the formulas
\begin{equation}\label{eq.ratios}
\frac{V_d}{V_u}=\frac{\rho_u}{\rho_d}
=\left(\frac{c_u}{c_d}\right)^2=\left(\frac{\xi_d}{\xi_u}\right)^2.
\end{equation}
The first equality in the above relations reflects the conservation of
the current. The second stems from the expression of the speed of
sound in the Gross-Pitaevskii approximation and the last one from the
fact that, still in the Gross-Pitaevskii framework,
$c_\alpha\,\xi_\alpha=\hbar/m$.

We have in both configurations
\begin{equation}\label{eq.wd2a}
\phi_d(x)=\exp(i\beta_d),
\end{equation}
where $\beta_d$ is a constant, indicating that the order parameter \eqref{eq.wd1} is a plane wave in the whole downstream
region $x\ge 0$. In the upstream region ($x\le 0$) we have instead
\begin{equation}\label{eq.wd2b}
\phi_u(x)=
\cos\theta \tanh \left(\frac{x-x_0}{\xi_u}\cos\theta\right)-i\sin\theta ,
\end{equation}
indicating that the upstream flow pattern corresponds to a fraction of a dark soliton, which can be assimilated to a plane wave 
only in the limit $x\to-\infty$. In \eqref{eq.wd2b}, $\theta\in[0,\pi/2]$ 
with $\sin\theta=M_u$.
\begin{figure}
    \centering
    \includegraphics[width=0.99\linewidth]{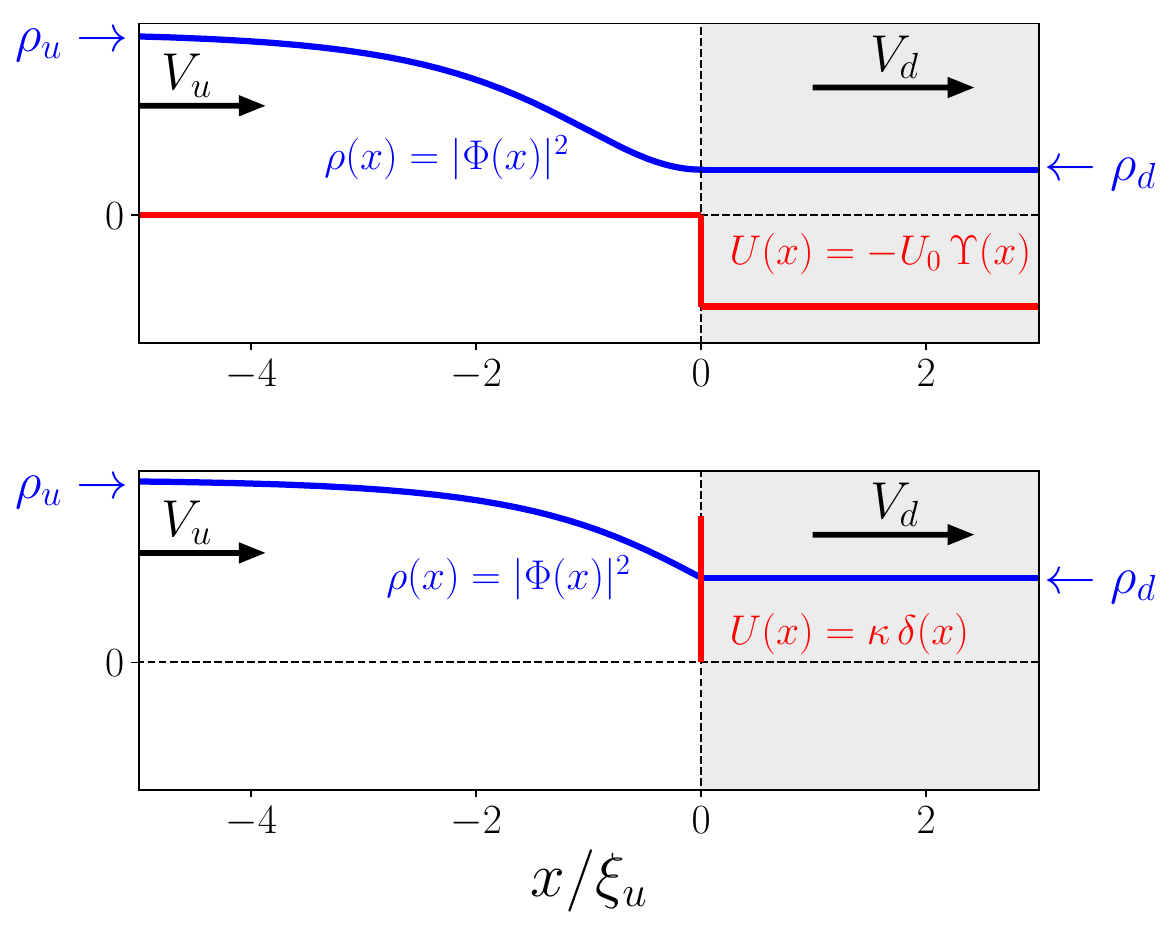}
    \caption{Upper plot: density profile of the waterfall
      configuration. Lower plot: density profile of the $\delta$-peak
      configuration. In both cases the order parameter takes the form
      of a plane wave downstream ($x>0$), and of a fraction of a gray
      soliton upstream ($x<0$). The downstream region is shaded to
      underline that is corresponds to the interior of the analog
      black hole. Note that, according to the convention used in the
      main text, all the quantities in this plot should be written
      with an index ``GP'', since they correspond to a (stationary)
      solution of Gross-Pitaevskii equation \eqref{eq16}. These
      indices are omitted here for legibility.}
    \label{fig:configs}
\end{figure}

After having listed properties valid in both configurations, let us
now consider the specifics of each of them.  For the waterfall
configuration the external potential in \eqref{eq16} is a step
function $U(x)=-U_0 \Upsilon(x)$, where $\Upsilon$ is the Heaviside
function. In this case, the order parameter defined by \eqref{eq.wd1},
\eqref{eq.wd2a} and \eqref{eq.wd2b} is a solution of the
Gross-Pitaevskii equation provided
\begin{equation}\label{eq.wd3}
x_0=0,\quad \beta_d=\pi,\quad
    \frac{U_0}{g\rho_u}=\frac{M_u^2}{2}+\frac{1}{2 M_u^2}-1,
\end{equation}
and
\begin{equation}\label{eq.wd4}
    \frac{V_d}{V_u}=\frac{1}{M_u^2}=M_d.
\end{equation}
The analog black hole realized in the 2019 Technion experiment \cite{deNova2019} is close to the waterfall configuration with $M_d=2.9$, cf. the discussion in \cite{Isoard2020}.

For the $\delta$-peak configuration, the external potential in \eqref{eq16} is a $\delta$ peak $U(x)=\kappa\, \delta(x)$. In this case, denoting as 
\begin{equation}\label{eq.defy}
y=-\tfrac12+\tfrac12 \sqrt{1+\frac{8}{M_u^2}},
\end{equation}
the order parameter \eqref{eq.wd1} is a solution of the Gross-Pitaevskii equation provided
\begin{equation}\label{eq.wd5}
\begin{split}
& \tanh\left(\frac{x_0}{\xi_u}\cos\theta\right)=\sqrt{\frac{y-1}{2}}\tan\theta
,\\
& \sin\beta_d=-M_u\sqrt{y},
\quad \kappa=\frac{\hbar^2}{m} \frac{M_u}{\xi_u}
\left(\frac{y-1}{2}\right)^{3/2},
\end{split}
\end{equation}
and
\begin{equation}\label{eq.wd6}
\frac{V_d}{V_u}
= \left(\frac{M_d}{M_u}\right)^{2/3}=y.
\end{equation}
The density profiles of the waterfall and $\delta$-peak configurations
are sketched in Fig. \ref{fig:configs}.  In both configurations
$M_u\le 1\le M_d$ and the knowledge of $M_u$ determines $M_d$ and all
the ratios between the upstream and downstream values of the relevant
parameters of the flow, as exposed in Eqs.  \eqref{eq.ratios},
\eqref{eq.wd4} and \eqref{eq.wd6}.

The flat profile configuration, introduced in
Refs. \cite{Balbinot2008,Carusotto2008} is a idealized setting in
which the Gross-Pitaevskii density and velocity are constant:
$\rho(x)=\rho_0$ and $V(x)=V_0$. An acoustic horizon can be
implemented in such a configuration by means of a step-like nonlinear
constant $g(x)$ combined with a step-like external potential $U(x)$:
\begin{equation}\label{eq.wd10}
g(x)=
\begin{cases}
    g_u & \mbox{when} \;\; x<0, \\
    g_d & \mbox{when} \quad x>0.
\end{cases}
\end{equation}
\begin{equation}\label{eq.wd11}
U(x)=
\begin{cases}
    U_u & \mbox{when} \;\; x<0, \\
    U_d & \mbox{when} \quad x>0.
\end{cases}
\end{equation}
The constancy of the density and of the velocity corresponds to a
background order parameter of the form \eqref{eq.wd1} with
$\rho_u=\rho_d=\rho_0$, $k_u=k_d=k_0=m V_0/\hbar$ and
$\phi_u(x)=\phi_d(x)=1$. The corresponding $\Phi(x)$ is made a
solution of the Gross-Pitaevskii equation by enforcing the relation
\begin{equation}\label{eq.wd12}
    g_u \rho_0 + U_u = g_d \rho_0 + U_d.
\end{equation}
In this configuration the whole of relation \eqref{eq.ratios} is not verified, but we have separately: 
\begin{equation}\label{eq.ratio_bis}
\frac{V_d}{V_u}=\frac{\rho_u}{\rho_d}=1 \quad\mbox{and}\quad
\frac{c_u}{c_d}=\frac{\xi_d}{\xi_u}=\frac{M_d}{M_u}.
\end{equation}
Note that, at variance with the waterfall and $\delta$-peak configurations, for the flat profile configuration the values of $M_u$ and $M_d$ are independent one from the other: any choice with $M_u<1$ and $M_d>1$ is acceptable.

\section{Elementary excitations and quantum modes}\label{app.modes}

\begin{figure}
    \centering
    \includegraphics[width=0.99\linewidth]{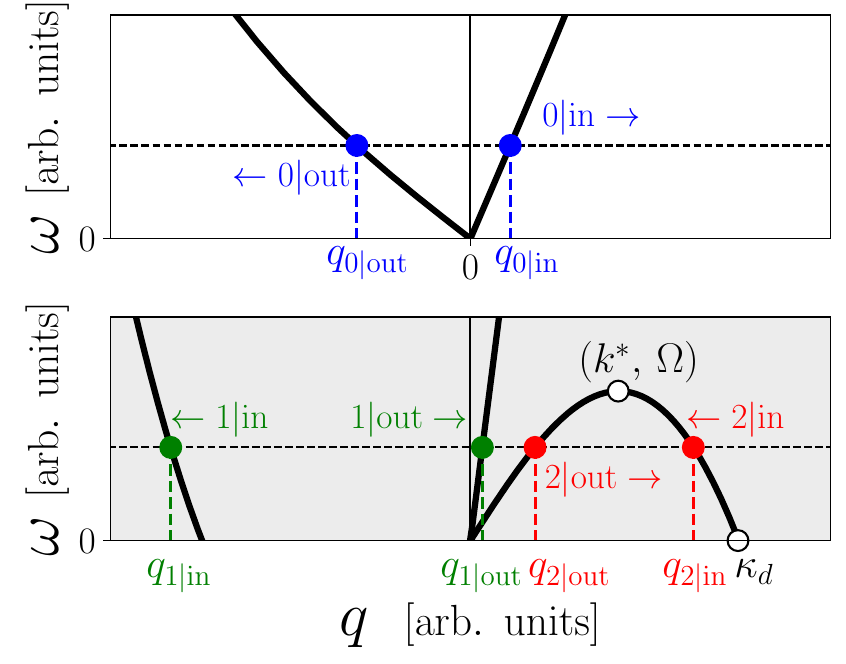}
\caption{Upper plot: dispersion relation \eqref{eq.m1} in the
  asymptotic upstream subsonic region. Lower plot: dispersion relation
  in the downstream supersonic region. This plot is shaded to
  emphasize that this region corresponds to the interior of the
  analog black hole.  The horizontal dashed line is the angular
  frequency $\omega$ of a given excitation. The colored points mark
  the corresponding excitation channels, of wavevectors $q_{0|{\rm
      out}}(\omega)$, $q_{0|{\rm in}}(\omega)$, $q_{1|{\rm
      in}}(\omega)$ etc. The arrows indicate the direction of
  propagation of the different channels. The wave vector $\kappa_d$
  corresponds to a zero energy channel, see
  Sec. \ref{sec.analog.stat.1D}.}
    \label{fig:dispersion}
\end{figure}

In a analog black configuration the flow is upstream subsonic and
downstream supersonic. The Doppler effect is thus different in these
regions, from which it stems that the asymptotic dispersion relations
are also different. In both asymptotic regions the flow is uniform
(with density $\rho_\alpha$ and velocity $V_\alpha$, where $\alpha=u$
in the upstream asymptotic region and $\alpha=d$ downstream, see
Appendix \ref{app.A}).  In these regions the elementary excitations
are thus plane waves.  Denoting their angular frequency as $\omega$
and their wave vector as $q$ we have\footnote{As in Appendix
\ref{app.A}, throughout this appendix we omit for legibility all the
subscripts ``GP'' and write $V_\alpha$ instead of $\Vgp^\alpha$ for
instance.}
\begin{equation}\label{eq.m1}
(\omega - q V_\alpha)^2 = \omega_{{\sss\rm B},\alpha}^2(q),
\end{equation}
where
\begin{equation}\label{eq.m2}
    \omega_{{\sss\rm B},\alpha}(q)=c_\alpha q \sqrt{1+q^2\xi_\alpha^2/4}
\end{equation}
is the usual Bogoliubov dispersion relation. In \eqref{eq.m2}
$c_\alpha$ and $\xi_\alpha$ are the speed of sound and healing length
in region $\alpha$. The dispersion relations \eqref{eq.m1} are
represented in Fig. \ref{fig:dispersion}. There are two upstream
propagation channels, which we denote as $0|{\rm in}$ and $0|{\rm
  out}$. The corresponding wave-vectors are denoted as $q_{0|{\rm
    in}}(\omega)$ and $q_{0|{\rm out}}(\omega)$. In the downstream
region, below an angular frequency which we denote as $\Omega$, there
are four propagation channels: $1|{\rm in}$, $1|{\rm out}$, $2|{\rm
  in}$ and $2|{\rm out}$, associated to wave vectors $q_{1|{\rm
    in}}(\omega)$, $q_{1|{\rm out}}(\omega)$, $q_{2|{\rm in}}(\omega)$
and $q_{2|{\rm out}}(\omega)$, respectively. Only $1|{\rm in}$ and
$1|{\rm out}$ survive when $\omega>\Omega$.
The propagation channels with note with a label ``in'' propagate towards
the horizon, the ones with label ``out'' propagate away from the
horizon.

Several propagation modes correspond to the channels identified in
Fig. \ref{fig:dispersion}.  For instance, a mode with we denote as
``ingoing'' and associate to a quantum operator $\hat{b}_0(\omega)$
corresponds to a wave incident in channel $0|{\rm in}$ at energy
$\hbar\omega$, scattered onto the horizon to the exit channels $0|{\rm
  out}$, $1|{\rm out}$ and $2|{\rm out}$ with amplitudes
$S_{0,0}(\omega)$, $S_{0,1}(\omega)$ and $S_{0,2}(\omega)$,
respectively. There are two other ingoing modes which are associated
to operators we denote as $\hat{b}_1(\omega)$ and
$\hat{b}_2(\omega)$. They correspond to modes initiated by a wave
incident in channel $1|{\rm in}$ and $2|{\rm in}$, respectively.
There are also three outgoing modes associated to scattering processes
resulting in the emission of a single wave along one of the three
“out” channels $0|{\rm out}$, $1|{\rm out}$, and $2|{\rm out}$. We
denote the corresponding quantum operators as $\hat{c}_0(\omega)$,
$\hat{c}_1(\omega)$ and $\hat{c}_2(\omega)$.  In short, the
modes, be they ingoing or outgoing, are identified by an index
$L\in\{0,1,2\}$ and the channels by an index $\ell\in\{0|{\rm
  in},0|{\rm out},1|{\rm in},1|{\rm out},2|{\rm in},2|{\rm out}\}$.
We also note that the outgoing modes $\hat{c}_0$ and $\hat{c}_2$ are
analogous to the modes denoted in the General Relativity context as
the Hawking and partner modes, respectively.

The outgoing modes relate to the incoming ones {\it via}
\begin{equation} \label{eq.m3}
\begin{pmatrix}
\hat{c}_{0}  \\
\hat{c}_{1} \\
\hat{c}_{2}^{\dagger} \\
\end{pmatrix}
=
\begin{pmatrix}
S_{00} & S_{01} & S_{02} \\
S_{10} & S_{11} & S_{12} \\
S_{20} & S_{21} & S_{22} \\
\end{pmatrix}
\,
\begin{pmatrix}
\hat{b}_{0}  \\
\hat{b}_{1} \\
\hat{b}_{2}^{\dagger} \\
\end{pmatrix},
\end{equation}
where for legibility we omit the $\omega$ dependence of all the
terms. The appearance of annihilation operators  $\hat{b}_2^\dagger$ and $\hat{c}_2^\dagger$ in
\eqref{eq.m3} reflects the fact that the modes $L=2$ have
a negative norm and should accordingly be quantized inverting the usual
role of the creation and annihilation operators  in order
that the elementary excitations satisfy the standard Bose commutation relations \cite{Blaizot1986}.

The $3\times 3$ scattering matrix $S(\omega)$ defined in \eqref{eq.m3} obeys a skew-unitarity relation \cite{Recati2009} :
\begin{equation}\label{eq.m4}
    S^\dagger \eta S =\eta =S \eta S^\dagger ,
    \quad\mbox{where}\quad
    \eta={\rm diag}(1,1,-1).
\end{equation}
For $\omega>\Omega$ the channels $2|{\rm in}$ and $2|{\rm out}$ disappear (cf.~Fig.~\ref{fig:dispersion}), as well as the modes $\hat{b}_2(\omega)$ and $\hat{c}_2(\omega)$. In this case 
the $S$-matrix becomes $2\times 2$ and unitary.

\section{Asymptotic source terms in a 1D black hole}\label{app.C}

The quantum fluctuation field $\hat{\psi}$ defined in \eqref{eq2} can
be expanded of the basis of ingoing modes defined in Appendix
\ref{app.modes} supplemented by the contribution of zero modes. As
explained in the main text, we will evaluate the source terms in
Eqs. \eqref{eqs4} and \eqref{eqs5} only in the asymptotic regions
($x\to\pm\infty$) where the contribution of the zero modes can be
omitted. In this case we have
\begin{equation}\label{eq.exp1}
\begin{split}
\hat{\psi}(x,t)= & e^{i k_\alpha x} 
\int_0^\infty \frac{d\omega}{\sqrt{2\pi}}
\sum_{L=0}^1 
\Big[ u_{\sss L}(x,\omega) e^{-i\omega t} \, \hat{b}_{\sss L}(\omega)\\
& +v_{\sss L}^*(x,\omega)e^{i\omega t} \, \hat{b}_{\sss L}^\dagger(\omega)\Big]\\
& +e^{i k_\alpha x} \int_0^\Omega \frac{d\omega}{\sqrt{2\pi}}
\Big[ u_{2}(x,\omega) e^{-i\omega t} \, \hat{b}^\dagger_{2}(\omega)\\
& +v_{2}^*(x,\omega)e^{i\omega t}\, \hat{b}_{2}(\omega)\Big],
\end{split}
\end{equation}
where\footnote{As in Appendixes \ref{app.A} and \ref{app.modes},
throughout this appendix we omit for legibility all the subscripts
``GP'' and write $V_\alpha$ instead of $\Vgp^\alpha$ for instance.}
$k_\alpha=k_u=m V_u/\hbar$ is $x<0$ and $k_\alpha=k_d=m V_d/\hbar$ if
$x>0$.  In this expression the $u_{\sss L}(x,\omega)$'s and $v_{\sss
  L}(x,\omega)$'s are linear combinations of the usual Bogoliubov
coefficients involving the coefficient of the scattering matrix
\eqref{eq.m3}, see Eqs, \eqref{eq.exp4} and \eqref{eq.exp5} below.

At zero temperature the different contributions to the source terms
which appear in Eqs. \eqref{eqs4} and \eqref{eqs5} without derivative
read \cite{Ciliberto2024_PhD}
\begin{equation}\label{eq.exp2}
\begin{aligned}
  g^{(2)}(x) =  & \dfrac{2}{\rho_\alpha \vert \phi_\alpha \vert^4}
  \int_0^\infty \dfrac{d\omega}{2\pi} \sum_{L=0}^1
  \Big[ \vert \tilde{v}_L \vert^2
    + \Re (\tilde{u}_{\sss L} \tilde{v}_{\sss L}^*) \Big] \\
  & \, +  \dfrac{2}{ \rho_\alpha \vert \phi_\alpha \vert^4}
  \int_0^\Omega \frac{d\omega}{2\pi} 
  \Big[ \vert \tilde{u}_{2} \vert^2 +
    \Re ( \tilde{u}_{2} \tilde{v}_{2}^*) \Big],
\end{aligned}
\end{equation}
and
\begin{equation}\label{eq.exp3}
\begin{aligned}
& \Re \langle \hat{\eta} \partial_{x}\hat{\theta} \rangle = 
\dfrac{1}{2\rho_\alpha |\phi_\alpha(x)|^2} \times
\\
& \int_0^\infty \dfrac{d\omega}{2\pi} \sum_{L=0}^2 \Im \left[ 
( \tilde{u}_{\sss L}^*+ \tilde{v}_{\sss L}^*) \partial_{x} \left(  
\dfrac{\tilde{u}_{\sss L}-\tilde{v}_{\sss L}}{|\phi_\alpha|^2}
\right) \right],
\end{aligned}
\end{equation}
where $\phi_\alpha(x)$ is defined in Appendix \ref{app.A},  $\tilde{u}_L(x,\omega)=u_L(x,\omega)\phi_\alpha^*(x)$ and $\tilde{v}_L(x,\omega)=v_L(x,\omega)\phi_\alpha(x)$, with $\alpha=u$ if $x<0$ 
and $\alpha=d$ if $x>0$.
In the asymptotic regions, the $\tilde{u}_{\sss L}$'s and 
$\tilde{v}_{\sss L}$'s are combinations of plane waves.
More precisely, deep in the upstream
subsonic region, i.e., when $x<0$, $x\ll -\xi_u$ :
\begin{eqnarray}\label{eq.exp4}
\begin{pmatrix}
\tilde{u}_{0} 
\\
\tilde{v}_{0}\end{pmatrix}
& = &
 S_{0,0}\, \begin{pmatrix}
\tilde{\cal U}_{0|{\rm out}}\\
\tilde{\cal V}_{0|{\rm out}}\end{pmatrix}
e^{{\rm i}q_{0|{\rm out}}x} +
\begin{pmatrix}
\tilde{\cal U}_{0|{\rm in}}\\
\tilde{\cal V}_{0|{\rm in}}\end{pmatrix}
e^{{\rm i}q_{0|{\rm in}}x},
\nonumber \\
\begin{pmatrix}{u}_{1}\\
{v}_{1}\end{pmatrix}
& = & S_{0,1}\, \begin{pmatrix}
\tilde{\cal U}_{0|{\rm out}}\\
\tilde{\cal V}_{0|{\rm out}}\end{pmatrix}
e^{{\rm i}q_{0|{\rm out}}x} , \\
\begin{pmatrix}
\tilde{u}_{2}\\
\tilde{v}_{2}\end{pmatrix}
& = & S_{0,2}\, \begin{pmatrix}
\tilde{\cal U}_{0|{\rm out}}\\
\tilde{\cal V}_{0|{\rm out}}\end{pmatrix}
e^{{\rm i}q_{0|{\rm out}}x} ,\nonumber
\end{eqnarray}
and deep in the downstream supersonic region, i.e., when $x>0$,
$x\gg \xi_d$ :
\begin{eqnarray}\label{eq.exp5}
\begin{pmatrix}
\tilde{u}_{0}\\
\tilde{v}_{0}\end{pmatrix}
&\!\! = \!\!& S_{1,0}  \begin{pmatrix}
\tilde{\cal U}_{d1|{\rm out}}\\
\tilde{\cal V}_{d1|{\rm out}}\end{pmatrix}
e^{{\rm i}q_{d1|{\rm out}}x} \nonumber \\
&\!\! + \!\!& S_{2,0} 
 \begin{pmatrix}
\tilde{\cal U}_{2|{\rm out}}\\
\tilde{\cal V}_{2|{\rm out}}
\end{pmatrix}
e^{{\rm i}q_{2|{\rm out}}x} , \nonumber \\
\begin{pmatrix}
\tilde{u}_{1}\\
\tilde{v}_{1}\end{pmatrix}
& \!\! = \!\!& 
S_{1,1} \begin{pmatrix} 
\tilde{\cal U}_{1|{\rm out}}\\
\tilde{\cal V}_{1|{\rm out}}\end{pmatrix}
e^{{\rm i}q_{1|{\rm out}}x} \\
& \!\! + \!\!&  S_{2,1} \begin{pmatrix}
\tilde{\cal U}_{2|{\rm out}}\\
\tilde{\cal V}_{2|{\rm out}}\end{pmatrix}
e^{{\rm i}q_{2|{\rm out}}x} 
+
\begin{pmatrix}
\tilde{\cal U}_{1|{\rm in}}\\
\tilde{\cal V}_{1|{\rm in}}\end{pmatrix} e^{{\rm i}q_{1|{\rm in}}x}
,\nonumber\\
\begin{pmatrix}
\tilde{u}_{2}\\
\tilde{v}_{2}\end{pmatrix}
& \!\! = \!\!& 
S_{1,2} \begin{pmatrix}
\tilde{\cal U}_{1|{\rm out}}\\
\tilde{\cal V}_{1|{\rm out}}\end{pmatrix}
e^{{\rm i}q_{1|{\rm out}}x} \nonumber \\
& \!\! + \!\!&  S_{2,2}
\begin{pmatrix}
\tilde{\cal U}_{2|{\rm out}}\\
\tilde{\cal V}_{2|{\rm out}}\end{pmatrix}
e^{{\rm i}q_{2|{\rm out}}x} 
+\begin{pmatrix}
\tilde{\cal U}_{2|{\rm in}}\\
\tilde{\cal V}_{2|{\rm in}}\end{pmatrix} e^{{\rm i}q_{2|{\rm in}}x}
.\nonumber
\end{eqnarray}
The Bogoliubov
coefficients $\tilde{\cal U}_\ell(\omega)$ and 
$\tilde{\cal V}_\ell(\omega)$ are real; their explicit expression is given in
Ref. \cite{Larre2012}. Note that the $q_\ell$'s and
$S_{i,j}$'s in
Eqs. \eqref{eq.exp2} and \eqref{eq.exp3} all depend on $\omega$, see Appendix \ref{app.modes}. 

From expressions \eqref{eq.exp2}, \eqref{eq.exp4} and \eqref{eq.exp5} we get the following expression of the asymptotic density-density correlation function $g_\alpha^{(2)}$ defined in \eqref{eqs6}:
\begin{equation}\label{eq.exp6}
g_u^{(2)}=-\frac{2}{\pi\xi_u\rho_u} + \frac{2}{\rho_u}
\int_0^\Omega \frac{d \omega}{2\pi} |S_{0,2}|^2 
\tilde{\cal R}_{0|{\rm out}}^2,
\end{equation}
\begin{equation}\label{eq.exp7}
\begin{split}
g_d^{(2)}= -\frac{2}{\pi\xi_d\rho_d} +
\frac{2}{\rho_d}
\int_0^\Omega \frac{d \omega}{2\pi}
 & \Big[
|S_{1,2}|^2 
\tilde{\cal R}_{1|{\rm out}}^2 +\\
& \left(|S_{2,2}|^2 -1\right)
\tilde{\cal R}_{2|{\rm out}}^2\Big]
,
\end{split}
\end{equation}
where $\tilde{\cal R}_\ell(\omega)=\tilde{\cal U}_\ell(\omega)+\tilde{\cal V}_\ell(\omega)$. Note that the skew unitarity \eqref{eq.m4} allows to write, in the last contribution to the integrand of \eqref{eq.exp7}:
$|S_{2,2}|^2 -1=|S_{2,0}|^2+|S_{2,1}|^2=|S_{0,2}|^2+|S_{1,2}|^2$. As a result, the Hawking contribution additional to the standard term $-2/\pi\xi_\alpha\rho_\alpha$ in $g_\alpha^{(2)}$ [cf. Eq. \eqref{eq.g21d}] is positive both in 
\eqref{eq.exp6} and \eqref{eq.exp7}.

According to \eqref{eq.exp6} and \eqref{eq.exp7} the explicit expression of the dimensionless quantities  ${\cal G}^{(H)}_\alpha$ defined in \eqref{eq12defa} is
\begin{equation}\label{eqB1}
    {\cal G}_u^{(H)} = \int_0^{\Omega_u} \!\frac{{\rm d}\varepsilon_u}{2\pi}
    |S_{0,2}|^2 N(|Q_{0|{\rm out}}|,M_u),
\end{equation}
and
\begin{equation}\label{eqB2}
\begin{split}
{\cal G}_d^{(H)} = \int_0^{\Omega_d}\!\frac{{\rm d}\varepsilon_d}{2\pi}
  \Big\{ & |S_{1,2}|^2 \, N(Q_{1|{\rm out}},-M_d)\\
- & (|S_{2,2}|^2-1) \, N(Q_{2|{\rm out}},M_d)
\Big\},
\end{split}
\end{equation}
where the $Q_\ell(\omega)$ are the dimensionless wave vectors: $Q_\ell=q_\ell \xi_u$ for $\ell=0|{\rm out}$ and $Q_\ell=q_\ell \xi_d$ for $\ell=1|{\rm out}$ 
and $\ell=2|{\rm out}$ (see the definition of the $q_\ell(\omega)$'s in 
Appendix \ref{app.modes}). We have also 
$\varepsilon_\alpha=\hbar\omega/(m c_\alpha^2)$, 
$\Omega_\alpha=\hbar\Omega/(m c_\alpha^2)$, where $\Omega$ is defined in Fig. \ref{fig:dispersion}, and
\begin{equation}\label{eqB3}
    N(Q,M)=\frac{Q}{1+Q^2/2-M\sqrt{1+Q^2/4}}.
\end{equation}
Similarly to what has just been done for the density-density
correlation function, we now describe the steps enabling to compute
the quantum contribution to the average current. From expressions
\eqref{eq.exp3}, \eqref{eq.exp4} and \eqref{eq.exp5} we get the
following expression for the asymptotic quantum contribution
$j_\alpha$ defined in \eqref{eqs6}:
\begin{equation}\label{eq.exp11}
j_u=\frac{\hbar}{m} \int_0^\Omega \frac{d\omega}{2\pi} \,
\frac{q_{0|{\rm out}}}{|\partial\omega/\partial q_{0|{\rm out}}|}
\,
|S_{0,2}|^2 ,
\end{equation}
and
\begin{equation}\label{eq.exp12}
\begin{split}
j_d=\frac{\hbar}{m}\int_0^\Omega \frac{d\omega}{2\pi} & \Big[
\frac{q_{1|{\rm out}}}{|\partial\omega/\partial q_{1|{\rm out}}|}
\, |S_{1,2}|^2\\
& -  
\frac{q_{2|{\rm out}}}{|\partial\omega/\partial q_{2|{\rm out}}|}
\, (|S_{2,2}|^2-1)\Big].
\end{split}
\end{equation}
We recall that $q_{0|{\rm out}}(\omega)<0$, whereas $q_{1|{\rm
    out}}(\omega)$ and $q_{2|{\rm out}}(\omega)$ are both positive,
see Fig. \ref{fig:dispersion}. Hence $j_u<0$ whereas the integrand of
$j_d$ contains two contributions, one positive and one negative. In
practice we find numerically that the second is dominant and that
$j_d$ is negative.  Accordingly to Eqs. \eqref{eq.exp11} and
\eqref{eq.exp12} the dimensionless quantities ${\cal J}^{(H)}_\alpha$
defined in Eq.  \eqref{eq12defb} read:
\begin{equation}\label{eqB4}
    {\cal J}_u^{(H)} = \int_0^{\Omega_u} \!\frac{{\rm d}\varepsilon_u}{2\pi}
    |S_{0,2}|^2 \, T(Q_{0|{\rm out}},M_u),
\end{equation}
and
\begin{equation}\label{eqB5}
\begin{split}
{\cal J}_d^{(H)} = \int_0^{\Omega_d} \!\frac{{\rm d}\varepsilon_d}{2\pi}
  \Big\{ & |S_{1,2}|^2 \, T(Q_{1|{\rm out}},-M_d)\\
+ & (|S_{2,2}|^2-1) \, T(Q_{2|{\rm out}},M_d)
\Big\},
\end{split}
\end{equation}
where 
\begin{equation}\label{eqB6}
    T(Q,M)=
    N(Q,M)\times \sqrt{1+Q^2/4}.
\end{equation}
We determine numerically the four quantities ${\cal G}_\alpha^{(H)}$
and ${\cal J}_\alpha^{(H)}$ ($\alpha=u$ and $d$) from expression
\eqref{eqB1}, \eqref{eqB2}, \eqref{eqB4} and \eqref{eqB5} after a
numerical computation of the elements of the $S$-matrix
\cite{Larre2012}.  It may be shown that these quantities cancel in the
limit $M_d \to 1$. This should be expected since in this limit the
negative norm channels $2|{\rm out}$ and $2|{\rm in}$ vanish,
resulting in a disappearance of Hawking radiation.  Also, since in
this limit the upper integration point in integrals \eqref{eqB1},
\eqref{eqB2}, \eqref{eqB3} and \eqref{eqB4} tend to zero, it is easy
to show that
\begin{equation}\label{eq,exp16}
{\cal J}_u^{(H)}\underset{M_d\to 1}{\simeq} - {\cal G}_u^{(H)}
\quad\mbox{and}\quad
{\cal J}_d^{(H)}\underset{M_d\to 1}{\simeq} - {\cal G}_d^{(H)}.
\end{equation}
The first equality follows directly from comparing expressions
\eqref{eqB1} and \eqref{eqB4} and noticing that $T(Q,M)\simeq N(Q,M)$
when $Q\to 0$ which is the appropriate limit to consider when
$\Omega\to 0$. The second one also follows from this property
complemented by the fact that, in the limit $M_d\to 1$, the second
terms of the integrands of \eqref{eqB2} and \eqref{eqB5} become
dominant.

It follows from these remarks that in the limit $M_d\to 1$ the 
leading terms of the series expansion 
of  ${\cal G}_\alpha^{(H)}$ and ${\cal J}_\alpha^{(H)}$ are of the form
\begin{subequations}\label{eqDL}
\begin{align}
{\cal G}_u^{(H)} & \simeq - {\cal J}_u^{(H)} 
\simeq {\cal H}_u (M_d-1),\\
{\cal G}_d^{(H)} & \simeq -{\cal J}_d^{(H)} 
\simeq {\cal H}_d (M_d-1),
\end{align}
\end{subequations}
where ${\cal H}_u$ and ${\cal H}_d$ are positive constants which we
determine numerically.  In the waterfall and $\delta$-peak
configurations $M_u$ also tends to 1 when $M_d$ does, and ${\cal H}_u$
and ${\cal H}_d$ are thus universal constants.  We find ${\cal
  H}_u=2.48\times 10^{-2}$ and ${\cal H}_d=9.36\times 10^{-3}$ in the
waterfall configuration, ${\cal H}_u=3.91\times 10^{-2}$ and ${\cal
  H}_d=5.34\times 10^{-2}$ in the $\delta$-peak configuration. On the
other hand, in the flat profile configuration ${\cal H}_u$ and ${\cal
  H}_d$ both depend on $M_u$.

\section{Derivation of Eq. \eqref{BR9}}\label{app.B}

In this appendix we make use of the Bogoliubov equations \eqref{eq.bogo} to cast the imaginary part of Eq. \eqref{BR6} under the form \eqref{BR9}.
From Eq. \eqref{eq.recast} this imaginary part reads
\begin{equation}\label{eq.B1}
\begin{split}
{\cal T} \left(\frac{\delta\rho}{2 \rho_{\rm\sss GP}}\right) + {\cal X} \delta \Theta = & 
-{\cal T} \Re\langle \hat{A}\rangle -{\cal X} \Im\langle \hat{A}\rangle\\
& +\frac{g \rho_{\rm\sss GP}}{2\hbar}
\langle \hat\eta\,\hat\theta + \hat\theta\,\hat\eta\rangle.
\end{split}
\end{equation}
To evaluate the source terms in the above we make use of the explicit
expression \eqref{eq39}. We first note that the action of ${\cal T}$
and ${\cal X}$ [defined in Eqs. \eqref{eq.ope}] on a product of
possibly non commuting fields $\hat{B}(\boldsymbol{x},t)$ and
$\hat{C}(\boldsymbol{x},t)$ reads
\begin{subequations}\label{eq.prod}
\begin{align}
& {\cal T} \hat{B}\hat{C}
= ({\cal T} \hat{B})\hat{C} + \hat{B} ({\cal T}\hat{C}),
\label{eq.proda}
\\
& {\cal X} \hat{B}\hat{C}
= ({\cal X} \hat{B})\hat{C} + \hat{B} ({\cal X}\hat{C})
+\frac{\hbar}{m} \boldsymbol{\nabla}\hat{B}\cdot
\boldsymbol{\nabla}\hat{C}
.
\label{eq.prodb}
\end{align}
\end{subequations}
Use of these relations and of Bogoliubov Eqs. \eqref{eq.bogo} yields
\begin{equation}\label{eq.B3}
{\cal T}\hat\theta^2 =
\tfrac12 \hat\theta({\cal X}\hat\eta) + 
\tfrac12 ({\cal X}\hat\eta)\hat\theta
-\frac{g \rho_{\rm\sss GP}}{\hbar}(\hat\theta\hat\eta+\hat\eta\hat\theta),
\end{equation}
\begin{equation}\label{eq.B4}
{\cal T}\hat\eta^2 =
-2 \hat\eta({\cal X}\hat\theta) - 2 ({\cal X}\hat\theta)\hat\eta,
\end{equation}
and
\begin{equation}\label{eq.B5}
{\cal T}(\hat\eta\hat\theta-\hat\theta\hat\eta)=0.
\end{equation}
In this last equation use has been made of the fact that the equal
time commutator of a quantum field and its spatial derivative cancels:
thus $\hat\eta({\cal X}\hat\eta)=({\cal X}\hat\eta)\hat\eta$ and
$\hat\theta({\cal X}\hat\theta)=({\cal X}\hat\theta)\hat\theta$.  We
also have
\begin{equation}\label{eq.B6}
\begin{split}
{\cal X}(\hat\theta\hat\eta+\hat\eta\hat\theta) = & 
({\cal X}\hat\theta)\hat\eta+\hat\theta({\cal X}\hat\eta) +
\frac{\hbar}{m}
\boldsymbol{\nabla}\hat\theta\cdot\boldsymbol{\nabla}\hat\eta
\\
 + &
({\cal X}\hat\eta)\hat\theta+\hat\eta({\cal X}\hat\theta) +
\frac{\hbar}{m}
\boldsymbol{\nabla}\hat\eta\cdot\boldsymbol{\nabla}\hat\theta.
\end{split}
\end{equation}
Combining the results \eqref{eq.B3}, \eqref{eq.B4}, \eqref{eq.B5} and
\eqref{eq.B6} enables us to express the source term in Eq. \eqref{eq.B1}
as:
\begin{equation}\label{eq.B7}
\begin{split}
&
-{\cal T} \Re\langle \hat{A}\rangle -{\cal X} \Im\langle \hat{A}\rangle
+\frac{g \rho_{\rm\sss GP}}{2\hbar}
\langle \hat\eta\,\hat\theta + \hat\theta\,\hat\eta\rangle=\\
& 
-\tfrac12 \left\langle ({\cal X}\hat\theta)\hat\eta
+ \hat\eta({\cal X}\hat\theta) \right\rangle 
\\
&
- \frac{\hbar}{4m}
\left\langle 
\boldsymbol{\nabla}\hat\theta\cdot\boldsymbol{\nabla}\hat\eta
+
\boldsymbol{\nabla}\hat\eta\cdot\boldsymbol{\nabla}\hat\theta
\right\rangle.
\end{split}
\end{equation}
From the definition \eqref{BR10} of $\hat{\boldsymbol{v}}$ and the
explicit expression \eqref{eq.opeb} we may write
\begin{equation}
{\cal X}\hat\theta=\frac{1}{2\rho_{\rm\sss GP}}
{\boldsymbol{\nabla}}\cdot(\rho_{\rm\sss GP}\hat{\boldsymbol{v}}),
\end{equation}
and thus Eq. \eqref{eq.B7} reads
\begin{equation}\label{eq.B9}
\begin{split}
& -{\cal T} \Re\langle \hat{A}\rangle -{\cal X} \Im\langle \hat{A}\rangle
+\frac{g \rho_{\rm\sss GP}}{2\hbar}
\langle \hat\eta\,\hat\theta + \hat\theta\,\hat\eta\rangle=\\
& -\frac{1}{4\rho_{\rm\sss GP}}
{\boldsymbol{\nabla}}\cdot
\langle\rho_{\rm\sss GP}\hat\eta\hat{\boldsymbol{v}} +
\rho_{\rm\sss GP}\hat{\boldsymbol{v}}\hat\eta\rangle
=\\
&
-\frac{1}{2\rho_{\rm\sss GP}}
{\boldsymbol{\nabla}}\cdot
\Re\langle\rho_{\rm\sss GP}\hat\eta\hat{\boldsymbol{v}}\rangle.
\end{split}
\end{equation}

The left hand side term of Eq. \eqref{eq.B1} can also be
simplified. To this end, let us first remark that, from current conservation
\begin{equation}\label{eq.B11}
\partial_t\rho_{\rm\sss GP}+\boldsymbol{\nabla}\cdot
(\rho_{\rm\sss GP}\boldsymbol{V}_{\!\!\sss GP})=0.
\end{equation}
From this relation and from the explicit definition \eqref{eq.opea} of
operator ${\cal T}$, it results that for any scalar quantity
$Y(\boldsymbol{x},t)$ we have
\begin{equation}\label{BR5}
{\cal T} \left(\frac{Y}{\rho_{\rm\sss GP}}\right)=
\frac{1}{\rho_{\rm\sss GP}}\partial_t Y + 
\frac{1}{\rho_{\rm\sss GP}}\boldsymbol{\nabla}\cdot (Y\, \boldsymbol{V}_{\!\!\sss GP}).
\end{equation}
Using relation \eqref{BR5} and the definition \eqref{BR10} of
$\delta\boldsymbol{V}$ makes it possible to write the left hand side
of Eq. \eqref{eq.B1} as
\begin{equation}\label{eq.B12}
\begin{split}
{\cal T} \left(\frac{\delta\rho}{2 \rho_{\rm\sss GP}}\right) 
+ {\cal X} \delta \Theta = &
\frac{1}{2\rho_{\rm\sss GP}} \partial_t\delta\rho +\\
& \frac{1}{2\rho_{\rm\sss GP}} 
\boldsymbol{\nabla}\cdot
\left(\delta\rho \boldsymbol{V}_{\!\!\sss GP} + \rho_{\rm\sss GP} \delta \boldsymbol{V}\right).
\end{split}
\end{equation}
Inserting expressions \eqref{eq.B9} and \eqref{eq.B12} in
\eqref{eq.B1} directly yields Eq. \eqref{BR9}.

\section{Asymptotic backreaction for $\delta$-peak and flat profile configurations}\label{app.E}
In the main text we present the modification of the asymptotic Mach
numbers for a waterfall configuration. The reason why we put the
emphasis on this configuration is that this is the only one which has
been realized experimentally so far \cite{deNova2019,Kolobov2021}. In
this appendix we consider two other configurations (dubbed
$\delta$-peak and flat profile in Appendix \ref{app.A}) in order to
check to what extent the waterfall configuration is typical.
\begin{figure}
    \centering
    \includegraphics[width=0.99\linewidth]{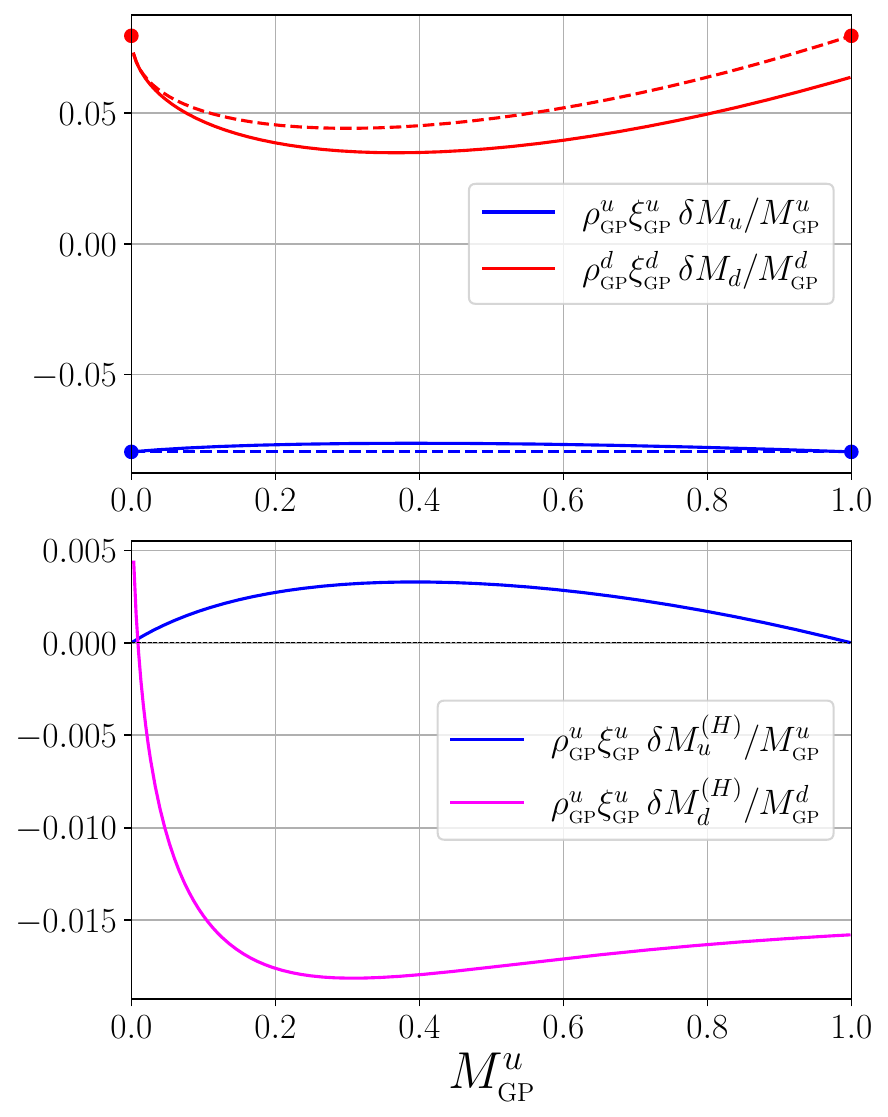}
    \caption{Same as Fig. \ref{fig:W} for the $\delta$-peak configuration.}
    \label{fig:DP}
\end{figure}

In Fig. \ref{fig:DP} we represent the modification of the upstream and
downstream Mach numbers due to quantum backreaction in a $\delta$-peak
configuration. The gross features of the results are the same as those
obtained for the waterfall configuration and presented in
Fig. \ref{fig:W}. A noticeable difference between the two figures is
that the part $\delta M_d^{(H)}$ of the downstream modifications
caused by Hawking radiation, while being roughly of the same order in
both settings, are positive for the waterfall configuration and
negative for the $\delta$-peak configuration (compare the lower panels
of Figs. \ref{fig:W} and \ref{fig:DP}).
\begin{figure}
    \centering
    \includegraphics[width=0.99\linewidth]{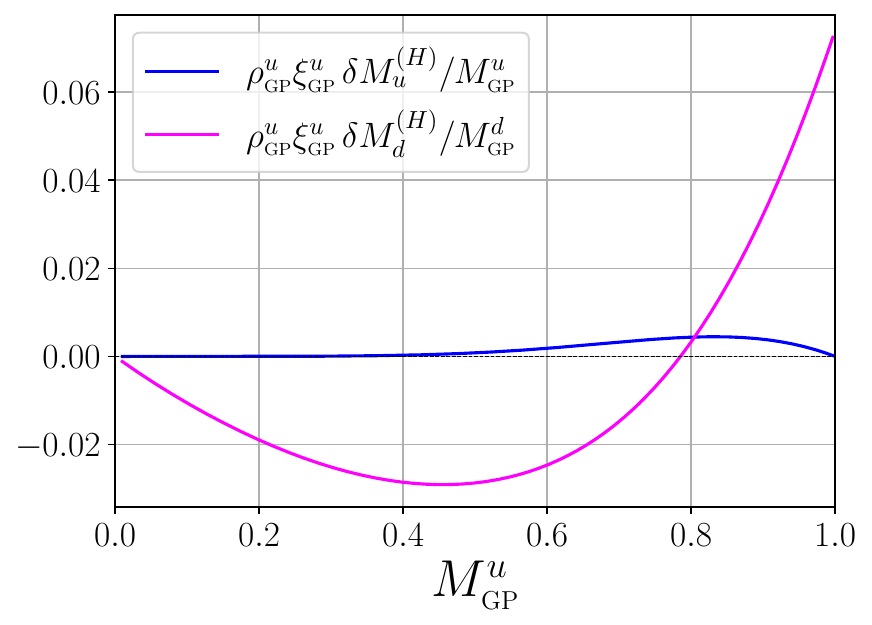}
    \caption{Hawking induced modifications $\delta M_\alpha^{(H)}$ of the asymptotic Mach numbers in a flat profile partially imitating a waterfall configuration.}
    \label{fig:FP_W}
\end{figure}
\begin{figure}
    \centering
    \includegraphics[width=0.99\linewidth]{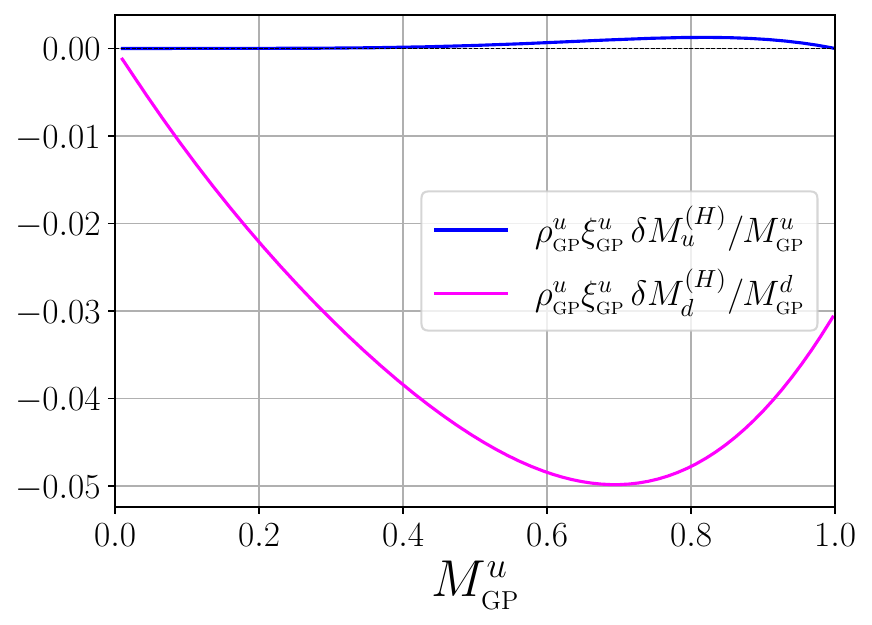}
    \caption{Same as Fig. \ref{fig:FP_W} in a flat profile partially imitating a $\delta$-peak configuration.}
    \label{fig:FP_DP}
\end{figure}
In order to assess if these different signs are signatures of specific
physical features we also determine $\delta M_\alpha^{(H)}$ for flat
profile configurations. As explained in Appendix \ref{app.A}, for
these configurations the values of the upstream and downstream Mach
numbers can be chosen independently one from the other. In order to
compare with the results of Figs. \ref{fig:W} and \ref{fig:DP} we
choose either a flat profile setting for which $\Mgp^u$ and $\Mgp^d$
are related through $\Mgp^d=(\Mgp^u)^{-2}$ [thus partially imitating a
  waterfall configuration, cf. Eq. \eqref{eq.wd4}], or through
$\Mgp^d/\Mgp^u= y^{3/2}$ where $y$ is defined in \eqref{eq.defy} [thus
  partially imitating a $\delta$-peak configuration,
  cf. Eq. \eqref{eq.wd6}]. The corresponding results are presented in
Figs. \ref{fig:FP_W} and \ref{fig:FP_DP}, respectively.

It appears that the Hawking contribution $\delta M_u^{(H)}$ to $\delta
M_u$ is positive in Figs. \ref{fig:W}, \ref{fig:DP}, \ref{fig:FP_W}
and \ref{fig:FP_DP}.  This can be understood from expression
\eqref{eq.dMsM} which implies that, for the boundary conditions
$\delta\mu=0=\delta V_u$, we have
\begin{equation}\label{eq.79}
  \frac{\delta M_u^{(H)}}{\Mgp^u}=
  -\tfrac12 \frac{\delta\rho_u^{(H)}}{\rhogp^u},
\end{equation}
where
\begin{equation}\label{eq.79bis}
    \delta\rho_u^{(H)}=-\frac{{\cal G}_u^{(H)}}{2\,\xigp^u} 
\end{equation}
is the part of $\delta\rho_u$ due to the Hawking backreaction. It is
shown in Appendix \ref{app.C} that ${\cal G}_\alpha^{(H)}$ is always
positive, which explains, {\it via} \eqref{eq.79} and
\eqref{eq.79bis}, why $\delta M_u^{(H)}$ is necessarily positive. In
contrast, no such simple relation exists in the downstream region,
where $\delta M_d^{(H)}$ can be either positive or negative.
Nevertheless, despite some differences, Figs. \ref{fig:DP},
\ref{fig:FP_W} and \ref{fig:FP_DP} support the discussion of the
waterfall configuration presented in the main text.

\end{appendix}

\pagebreak

\bibliography{biblio}

@article{Bogoliubov1947,
    author = "Bogoliubov, N. N.",
    title = "{On the theory of superfluidity}",
    journal = "J. Phys. (USSR)",
    volume = "11",
    pages = "23--32",
    year = "1947"
}

@article{Lee1957,
  title = {Eigenvalues and Eigenfunctions of a {Bose} System of Hard Spheres and Its Low-Temperature Properties},
  author = {Lee, T. D. and Huang, Kerson and Yang, C. N.},
  journal = {Phys. Rev.},
  volume = {106},
  issue = {6},
  pages = {1135--1145},
  numpages = {0},
  year = {1957},
  month = {Jun},
  publisher = {American Physical Society},
  doi = {10.1103/PhysRev.106.1135},
  url = {https://link.aps.org/doi/10.1103/PhysRev.106.1135}
}

@article{Beliaev1958a,
author={Beliaev, S. T.},
title={Application of the methods of quantum field theory to a system of bosons},
journal={Sov. Phys. JETP},
volume={34},
pages={289},
year={1958}
}

@article{Beliaev1958b,
author={Beliaev, S. T.},
title={Energy spectrum of a non-ideal {Bose} gas},
journal={Sov. Phys. JETP},
volume={34},
pages={299},
year={1958}
}

@article{Hugenholtz1959,
  title = {Ground-State Energy and Excitation Spectrum of a System of Interacting Bosons},
  author = {Hugenholtz, N. M. and Pines, D.},
  journal = {Phys. Rev.},
  volume = {116},
  issue = {3},
  pages = {489--506},
  numpages = {0},
  year = {1959},
  month = {Nov},
  publisher = {American Physical Society},
  doi = {10.1103/PhysRev.116.489},
  url = {https://link.aps.org/doi/10.1103/PhysRev.116.489}
}

@article{Lieb1963a,
  title = {Exact Analysis of an Interacting {Bose} Gas. {I}. {The} General Solution and the Ground State},
  author = {Lieb, Elliott H. and Liniger, Werner},
  journal = {Phys. Rev.},
  volume = {130},
  issue = {4},
  pages = {1605--1616},
  numpages = {0},
  year = {1963},
  month = {May},
  publisher = {American Physical Society},
  doi = {10.1103/PhysRev.130.1605},
  url = {https://link.aps.org/doi/10.1103/PhysRev.130.1605}
}

@article{Lieb1963b,
  title = {Exact Analysis of an Interacting {Bose} Gas. {II}. {The} Excitation Spectrum},
  author = {Lieb, Elliott H.},
  journal = {Phys. Rev.},
  volume = {130},
  issue = {4},
  pages = {1616--1624},
  numpages = {0},
  year = {1963},
  month = {May},
  publisher = {American Physical Society},
  doi = {10.1103/PhysRev.130.1616},
  url = {https://link.aps.org/doi/10.1103/PhysRev.130.1616}
}

@article{Hohenberg1965,
title = {Microscopic theory of superfluid helium},
author = {P.C Hohenberg and P.C Martin},
journal = {Ann. Phys. (N. Y.)},
volume = {34},
number = {2},
pages = {291-359},
year = {1965},
issn = {0003-4916},
doi = {https://doi.org/10.1016/0003-4916(65)90280-0}
}

@article{Anderson1966,
  title = {Considerations on the Flow of Superfluid Helium},
  author = {Anderson, P. W.},
  journal = {Rev. Mod. Phys.},
  volume = {38},
  issue = {2},
  pages = {298--310},
  numpages = {0},
  year = {1966},
  month = {Apr},
  publisher = {American Physical Society},
  doi = {10.1103/RevModPhys.38.298},
  url = {https://link.aps.org/doi/10.1103/RevModPhys.38.298}
}

@article{Popov1971,
title={Application of functional integration to the derivation of the low-frequency asymptotic behaviour of {Green's} functions and kinetic equations for a nonideal {Bose} gas},
author={Popov, V. N.},
journal={Theor. Math. Phys.},
volume={6},
pages={65},
year={1971},
doi={10.1007/BF01037581}
}

@article{Popov1972a,
title={Hydrodynamic Hamiltonian for a nonideal {Bose} gas},
author={Popov, V. N.},
journal={Theor. Math. Phys.},
volume={11},
pages={478},
year={1972},
doi={10.1007/BF01028563}
}

@article{Popov1972b,
title={On the theory of the superfluidity of two- and one-dimensional {Bose} systems},
author={Popov, V. N.},
journal={Theor. Math. Phys.},
volume={11},
pages={565},
year={1972},
doi={10.1007/BF01028373}
}

@article{Unruh1981,
  title = {Experimental Black-Hole Evaporation?},
  author = {Unruh, W. G.},
  journal = {Phys. Rev. Lett.},
  volume = {46},
  issue = {21},
  pages = {1351--1353},
  numpages = {0},
  year = {1981},
  month = {May},
  publisher = {American Physical Society},
  doi = {10.1103/PhysRevLett.46.1351},
  url = {https://link.aps.org/doi/10.1103/PhysRevLett.46.1351}
}

@article{Bardeen1981,
  title = {Black Holes Do Evaporate Thermally},
  author = {Bardeen, James M.},
  journal = {Phys. Rev. Lett.},
  volume = {46},
  issue = {6},
  pages = {382--385},
  numpages = {0},
  year = {1981},
  month = {Feb},
  publisher = {American Physical Society},
  doi = {10.1103/PhysRevLett.46.382},
  url = {https://link.aps.org/doi/10.1103/PhysRevLett.46.382}
}

@incollection{York1984,
author = {York, J.W. Jr.},
title  = {What happens to the horizon when a black hole radiates},
booktitle = {Quantum theory of gravity. Essays in honor of the 60th
birthday of {Bryce S. DeWitt}},
editor={Steven M. Christensen},
publisher    = {Adam Hilger},
year = {1984},
address = {Bristol},
pages= {135-147}  
}

@article{Balbinot1986,
  title = {Back reaction and the small-mass regime},
  author = {Balbinot, Roberto},
  journal = {Phys. Rev. D},
  volume = {33},
  issue = {6},
  pages = {1611--1615},
  numpages = {0},
  year = {1986},
  month = {Mar},
  publisher = {American Physical Society},
  doi = {10.1103/PhysRevD.33.1611},
  url = {https://link.aps.org/doi/10.1103/PhysRevD.33.1611}
}

@book{Blaizot1986,
  title = {Quantum Theory of Finite Systems},
  author = {Blaizot, Jean-Paul and Ripka, Georges},
  year = {1986},
  publisher = {{MIT Press}},
  address = {{Cambridge, Mass}}
}

@article{Shevchenko1992,
    author = {Shevchenko, S. I.},
    title = {On the theory of a {Bose} gas in a nonuniform field},
    journal = {Sov. J. Low Temp. Phys.},
    volume = {18},
    number = {4},
    pages = {223-230},
    year = {1992},
    issn = {0360-0335},
    doi = {10.1063/10.0033126}
}

@article{Griffin1996,
  title = {Conserving and gapless approximations for an inhomogeneous {Bose} gas at finite temperatures},
  author = {Griffin, A.},
  journal = {Phys. Rev. B},
  volume = {53},
  issue = {14},
  pages = {9341--9347},
  numpages = {0},
  year = {1996},
  month = {Apr},
  publisher = {American Physical Society},
  doi = {10.1103/PhysRevB.53.9341},
  url = {https://link.aps.org/doi/10.1103/PhysRevB.53.9341}
}

@article{Giorgini1998,
  title = {Damping in dilute {Bose} gases: A mean-field approach},
  author = {Giorgini, S.},
  journal = {Phys. Rev. A},
  volume = {57},
  issue = {4},
  pages = {2949--2957},
  numpages = {0},
  year = {1998},
  month = {Apr},
  publisher = {American Physical Society},
  doi = {10.1103/PhysRevA.57.2949},
  url = {https://link.aps.org/doi/10.1103/PhysRevA.57.2949}
}

@article{Fedichev1998,
  title = {Finite-temperature perturbation theory for a spatially inhomogeneous {Bose-condensed} gas},
  author = {Fedichev, P. O. and Shlyapnikov, G. V.},
  journal = {Phys. Rev. A},
  volume = {58},
  issue = {4},
  pages = {3146--3158},
  numpages = {0},
  year = {1998},
  month = {Oct},
  publisher = {American Physical Society},
  doi = {10.1103/PhysRevA.58.3146},
  url = {https://link.aps.org/doi/10.1103/PhysRevA.58.3146}
}

@article{Shi1998,
title = {Finite-temperature excitations in a dilute {Bose-condensed} gas},
journal = {Phys. Rep.},
volume = {304},
number = {1},
pages = {1-87},
year = {1998},
issn = {0370-1573},
doi = {https://doi.org/10.1016/S0370-1573(98)00015-5},
author = {Hua Shi and Allan Griffin}
}

@article{Zaremba1999,
title={Dynamics of Trapped {Bose} Gases at Finite Temperatures},
author={Zaremba, E. and Nikuni, T. and Griffin, A.},
journal={J. Low Temp. Phys.},
volume={116},
pages={277},
year={1999},
doi={10.1023/A:1021846002995}
}

@article{Giorgini2000,
  title = {Collisionless dynamics of dilute {Bose} gases: {Role} of quantum and thermal fluctuations},
  author = {Giorgini, S.},
  journal = {Phys. Rev. A},
  volume = {61},
  issue = {6},
  pages = {063615},
  numpages = {23},
  year = {2000},
  month = {May},
  publisher = {American Physical Society},
  doi = {10.1103/PhysRevA.61.063615},
  url = {https://link.aps.org/doi/10.1103/PhysRevA.61.063615}
}

@article{Garay2000,
  title = {Sonic Analog of Gravitational Black Holes in {Bose-Einstein} Condensates},
  author = {Garay, L. J. and Anglin, J. R. and Cirac, J. I. and Zoller, P.},
  journal = {Phys. Rev. Lett.},
  volume = {85},
  issue = {22},
  pages = {4643--4647},
  numpages = {0},
  year = {2000},
  month = {Nov},
  publisher = {American Physical Society},
  doi = {10.1103/PhysRevLett.85.4643},
  url = {https://link.aps.org/doi/10.1103/PhysRevLett.85.4643}
}

@article{Petrov2000,
  title = {Regimes of Quantum Degeneracy in Trapped {1D} Gases},
  author = {Petrov, D. S. and Shlyapnikov, G. V. and Walraven, J. T. M.},
  journal = {Phys. Rev. Lett.},
  volume = {85},
  issue = {18},
  pages = {3745--3749},
  numpages = {0},
  year = {2000},
  month = {Oct},
  publisher = {American Physical Society},
  doi = {10.1103/PhysRevLett.85.3745},
  url = {https://link.aps.org/doi/10.1103/PhysRevLett.85.3745}
}

@article{Andersen2002,
  title = {Phase Fluctuations in Atomic {Bose} Gases},
  author = {Andersen, J. O. and Al Khawaja, U. and Stoof, H. T. C.},
  journal = {Phys. Rev. Lett.},
  volume = {88},
  issue = {7},
  pages = {070407},
  numpages = {4},
  year = {2002},
  month = {Feb},
  publisher = {American Physical Society},
  doi = {10.1103/PhysRevLett.88.070407},
  url = {https://link.aps.org/doi/10.1103/PhysRevLett.88.070407}
}

@article{Gangardt2003,
  title = {Stability and Phase Coherence of Trapped {1D Bose} Gases},
  author = {Gangardt, D. M. and Shlyapnikov, G. V.},
  journal = {Phys. Rev. Lett.},
  volume = {90},
  issue = {1},
  pages = {010401},
  numpages = {4},
  year = {2003},
  month = {Jan},
  publisher = {American Physical Society},
  doi = {10.1103/PhysRevLett.90.010401},
  url = {https://link.aps.org/doi/10.1103/PhysRevLett.90.010401}
}

@article{Mora2003,
  title = {Extension of {Bogoliubov} theory to quasicondensates},
  author = {Mora, Christophe and Castin, Yvan},
  journal = {Phys. Rev. A},
  volume = {67},
  issue = {5},
  pages = {053615},
  numpages = {24},
  year = {2003},
  month = {May},
  publisher = {American Physical Society},
  doi = {10.1103/PhysRevA.67.053615},
  url = {https://link.aps.org/doi/10.1103/PhysRevA.67.053615}
}

@article{Petrov2004,
title={Low-dimensional trapped gases},
author={Petrov, D. S. and Gangardt, D. M. and Shlyapnikov, G. V.},
journal={J. Phys. IV France},
volume={116},
year={2004},
pages={5-44},
doi={https://doi.org/10.1051/jp4:2004116001}
}

@article{Balbinot2005a,
  title = {Backreaction in Acoustic Black Holes},
  author = {Balbinot, Roberto and Fagnocchi, Serena and Fabbri, Alessandro and Procopio, Giovanni P.},
  journal = {Phys. Rev. Lett.},
  volume = {94},
  issue = {16},
  pages = {161302},
  numpages = {4},
  year = {2005},
  month = {Apr},
  publisher = {American Physical Society},
  doi = {10.1103/PhysRevLett.94.161302},
  url = {https://link.aps.org/doi/10.1103/PhysRevLett.94.161302}
}

@article{Balbinot2005,
  title = {Quantum effects in acoustic black holes: The backreaction},
  author = {Balbinot, R. and Fagnocchi, S. and Fabbri, A.},
  journal = {Phys. Rev. D},
  volume = {71},
  issue = {6},
  pages = {064019},
  numpages = {11},
  year = {2005},
  month = {Mar},
  publisher = {American Physical Society},
  doi = {10.1103/PhysRevD.71.064019},
  url = {https://link.aps.org/doi/10.1103/PhysRevD.71.064019}
}

@book{Fabbri_2005,
	title = {Modeling {Black} {Hole} {Evaporation}},
	isbn = {9781860945274 9781860947223},
	url = {http://www.worldscientific.com/worldscibooks/10.1142/p378},
	urldate = {2021-06-07},
	publisher = {Imperial College Press},
	author = {Fabbri, Alessandro and Navarro-Salas, Jos\'e},
	month = jan,
	year = {2005},
	doi = {10.1142/p378},
}

@article{Schultzhold2005,
  title = {Quantum backreaction in dilute {Bose-Einstein} condensates},
  author = {Sch\"utzhold, Ralf and Uhlmann, Michael and Xu, Yan and Fischer, Uwe R.},
  journal = {Phys. Rev. D},
  volume = {72},
  issue = {10},
  pages = {105005},
  numpages = {8},
  year = {2005},
  month = {Nov},
  publisher = {American Physical Society},
  doi = {10.1103/PhysRevD.72.105005},
  url = {https://link.aps.org/doi/10.1103/PhysRevD.72.105005}
}

@article{Hu2007,
  title = {Metric fluctuations of an evaporating black hole from backreaction of stress tensor fluctuations},
  author = {Hu, B. L. and Roura, Albert},
  journal = {Phys. Rev. D},
  volume = {76},
  issue = {12},
  pages = {124018},
  numpages = {19},
  year = {2007},
  month = {Dec},
  publisher = {American Physical Society},
  doi = {10.1103/PhysRevD.76.124018},
  url = {https://link.aps.org/doi/10.1103/PhysRevD.76.124018}
}

@article{Balbinot2008,
 title = {Nonlocal density correlations as a signature of {Hawking} radiation from acoustic black holes},
 author = {Balbinot, Roberto and Fabbri, Alessandro and Fagnocchi, Serena and Recati, Alessio and Carusotto, Iacopo},
 journal = {Phys. Rev. A},
volume = {78},
issue = {2},
  pages = {021603},
  numpages = {4},
  year = {2008},
  month = {Aug},
  publisher = {American Physical Society},
  doi = {10.1103/PhysRevA.78.021603},
  url = {https://link.aps.org/doi/10.1103/PhysRevA.78.021603}
}

@article{Carusotto2008,
doi = {10.1088/1367-2630/10/10/103001},
url = {https://dx.doi.org/10.1088/1367-2630/10/10/103001},
year = {2008},
month = {oct},
publisher = {},
volume = {10},
number = {10},
pages = {103001},
author = {Carusotto, Iacopo and Fagnocchi, Serena and Recati, Alessio and Balbinot, Roberto and Fabbri, Alessandro},
title = {Numerical observation of {Hawking} radiation from acoustic black holes in atomic {Bose–Einstein} condensates},
journal = {New J. Phys.}
}

@book{GNZ2009, 
place={Cambridge}, 
booktitle={Bose-Condensed Gases at Finite Temperatures}, 
publisher={Cambridge University Press}, 
author={Griffin, Allan and Nikuni, Tetsuro and Zaremba, Eugene}, 
year={2009},
doi={10.1017/CBO9780511575150}
}

@article{Recati2009,
  title = {{Bogoliubov} theory of acoustic {Hawking} radiation in
                  {Bose-Einstein} condensates},
  author = {Recati, A. and Pavloff, N. and Carusotto, I.},
  journal = {Phys. Rev. A},
  volume = {80},
  issue = {4},
  pages = {043603},
  numpages = {10},
  year = {2009},
  month = {Oct},
  publisher = {American Physical Society},
  doi = {10.1103/PhysRevA.80.043603},
  url = {https://link.aps.org/doi/10.1103/PhysRevA.80.043603}
}

@article{Mayoral2011,
doi = {10.1088/1367-2630/13/2/025007},
url = {https://dx.doi.org/10.1088/1367-2630/13/2/025007},
year = {2011},
month = {feb},
publisher = {},
volume = {13},
number = {2},
pages = {025007},
author = {Mayoral, Carlos and Recati, Alessio and Fabbri, Alessandro and Parentani, Renaud and Balbinot, Roberto and Carusotto, Iacopo},
title = {Acoustic white holes in flowing atomic {Bose–Einstein} condensates},
journal = {New J. Phys.}
}

@book{Pethick2011,
  title = {Bose-{{Einstein}} Condensation in dilute gases},
  author = {Pethick, C. J. and Smith, H.},
  year = {2011},
  publisher = {{Cambridge University Press}},
  address = {{Cambridge, United Kingdom}},
  series = {International Series of Monographs on Physics}
}

@article{Kamchatnov2012,
title = {Generation of dispersive shock waves by the flow of a {Bose-Einstein} condensate past a narrow obstacle},
author = {Kamchatnov, A. M. and Pavloff, N.},
journal = {Phys. Rev. A},
volume = {85},
issue = {3},
pages = {033603},
numpages = {11},
year = {2012},
month = {Mar},
publisher = {American Physical Society},
doi = {10.1103/PhysRevA.85.033603},
url = {https://link.aps.org/doi/10.1103/PhysRevA.85.033603}
}

@article{Larre2012,
  title = {Quantum fluctuations around black hole horizons in {Bose-Einstein} condensates},
  author = {Larr\'e, P.-\'E. and Recati, A. and Carusotto, I. and Pavloff, N.},
  journal = {Phys. Rev. A},
  volume = {85},
  issue = {1},
  pages = {013621},
  numpages = {16},
  year = {2012},
  month = {Jan},
  publisher = {American Physical Society},
  doi = {10.1103/PhysRevA.85.013621},
  url = {https://link.aps.org/doi/10.1103/PhysRevA.85.013621}
}

@article{Chiocchetta2013,
doi = {10.1209/0295-5075/102/67007},
url = {https://dx.doi.org/10.1209/0295-5075/102/67007},
year = {2013},
month = {jul},
publisher = {EDP Sciences, IOP Publishing and Società Italiana di Fisica},
volume = {102},
number = {6},
pages = {67007},
author = {Chiocchetta, A. and Carusotto, I.},
title = {Non-equilibrium quasi-condensates in reduced dimensions},
journal = {Europhys. Lett.}
}

@article{Ji2015,
  title = {Temporal coherence of one-dimensional nonequilibrium quantum fluids},
  author = {Ji, Kai and Gladilin, Vladimir N. and Wouters, Michiel},
  journal = {Phys. Rev. B},
  volume = {91},
  issue = {4},
  pages = {045301},
  numpages = {6},
  year = {2015},
  month = {Jan},
  publisher = {American Physical Society},
  doi = {10.1103/PhysRevB.91.045301},
  url = {https://link.aps.org/doi/10.1103/PhysRevB.91.045301}
}

@article{Michel2015,
  title = {Nonlinear effects in time-dependent transonic flows: An analysis of analog black hole stability},
  author = {Michel, Florent and Parentani, Renaud},
  journal = {Phys. Rev. A},
  volume = {91},
  issue = {5},
  pages = {053603},
  numpages = {20},
  year = {2015},
  month = {May},
  publisher = {American Physical Society},
  doi = {10.1103/PhysRevA.91.053603},
  url = {https://link.aps.org/doi/10.1103/PhysRevA.91.053603}
}

@article{Anderson2015,
  title = {Low frequency gray-body factors and infrared divergences: Rigorous results},
  author = {Anderson, Paul R. and Fabbri, Alessandro and Balbinot, Roberto},
  journal = {Phys. Rev. D},
  volume = {91},
  issue = {6},
  pages = {064061},
  numpages = {18},
  year = {2015},
  month = {Mar},
  publisher = {American Physical Society},
  doi = {10.1103/PhysRevD.91.064061},
  url = {https://link.aps.org/doi/10.1103/PhysRevD.91.064061}
}

@book{Pitaevskii2016,
  title = {Bose-{{Einstein}} Condensation and Superfluidity},
  author = {Pitaevskii, L. P. and Stringari, S.},
  year = {2016},
  publisher = {{Oxford University Press}},
  address = {{Oxford, United Kingdom}},
  series = {International Series of Monographs on Physics}
}

@article{Steinhauer2016,
	title = {Observation of quantum {Hawking} radiation and its entanglement in an analogue black hole},
	volume = {12},
	copyright = {2016 Nature Publishing Group},
	issn = {1745-2481},
	url = {https://www.nature.com/articles/nphys3863},
	doi = {10.1038/nphys3863},
	number = {10},
	urldate = {2020-10-01},
	journal = {Nat. Phys.},
	author = {Steinhauer, Jeff},
	month = oct,
	year = {2016},
	pages = {959--965}
}

@article{Wang2017,
	title={Induced density correlations in a sonic black hole condensate},
	author={Yi-Hsieh Wang and Ted Jacobson and Mark Edwards and Charles W. Clark},
	journal={SciPost Phys.},
	volume={3},
	pages={022},
	year={2017},
	publisher={SciPost},
	doi={10.21468/SciPostPhys.3.3.022},
	url={https://scipost.org/10.21468/SciPostPhys.3.3.022},
}

@article{Robertson2018,
  title = {Nonlinearities induced by parametric resonance in effectively {1D} atomic {Bose} condensates},
  author = {Robertson, Scott and Michel, Florent and Parentani, Renaud},
  journal = {Phys. Rev. D},
  volume = {98},
  issue = {5},
  pages = {056003},
  numpages = {24},
  year = {2018},
  month = {Sep},
  publisher = {American Physical Society},
  doi = {10.1103/PhysRevD.98.056003},
  url = {https://link.aps.org/doi/10.1103/PhysRevD.98.056003}
}

@article{Chatrchyan2018,
  title = {Analog cosmological reheating in an ultracold {Bose} gas},
  author = {Chatrchyan, Aleksandr and Geier, Kevin T. and Oberthaler, Markus K. and Berges, J\"urgen and Hauke, Philipp},
  journal = {Phys. Rev. A},
  volume = {104},
  issue = {2},
  pages = {023302},
  numpages = {27},
  year = {2021},
  month = {Aug},
  publisher = {American Physical Society},
  doi = {10.1103/PhysRevA.104.023302},
  url = {https://link.aps.org/doi/10.1103/PhysRevA.104.023302}
}

@article{deNova2019,
title = {Observation of thermal {Hawking} radiation and its temperature in an analogue black hole},
volume = {569},
doi = {10.1038/s41586-019-1241-0},
number = {7758},
urldate = {2020-10-01},
journal = {Nature (London)},
author = {Muñoz de Nova, J. R. and Golubkov, K. and Kolobov, V. I. and Steinhauer, Jeff},
month = may,
year = {2019},
pages = {688--691}
}

@article{Butera2019,
doi = {10.1209/0295-5075/128/24002},
url = {https://dx.doi.org/10.1209/0295-5075/128/24002},
year = {2020},
month = {jan},
publisher = {EDP Sciences, IOP Publishing and Società Italiana di Fisica},
volume = {128},
number = {2},
pages = {24002},
author = {Butera, Salvatore and Carusotto, Iacopo},
title = {Quantum fluctuations of the friction force induced by the dynamical {Casimir} emission},
journal = {Europhys. Lett.}
}

@Article{Liberati2020,
AUTHOR = {Liberati, Stefano and Tricella, Giovanni and Trombettoni, Andrea},
TITLE = {Back-Reaction in Canonical Analogue Black Holes},
JOURNAL = {Appl. Sci.},
VOLUME = {10},
YEAR = {2020},
NUMBER = {24},
pages = {8868},
URL = {https://www.mdpi.com/2076-3417/10/24/8868},
ISSN = {2076-3417},
DOI = {10.3390/app10248868}
}

@article{Isoard2020,
  title = {Departing from Thermality of Analogue {Hawking} Radiation in a {Bose-Einstein} Condensate},
  author = {Isoard, M. and Pavloff, N.},
  journal = {Phys. Rev. Lett.},
  volume = {124},
  issue = {6},
  pages = {060401},
  numpages = {6},
  year = {2020},
  month = {Feb},
  publisher = {American Physical Society},
  doi = {10.1103/PhysRevLett.124.060401},
  url = {https://link.aps.org/doi/10.1103/PhysRevLett.124.060401}
}

@article{Tettamanti2020,
author = {Tettamanti, Manuele and Parola, Alberto},
title = {Formation dynamics of black- and white-hole horizons in an analogue gravity model},
doi = {10.3390/universe6080105},
journal = {Universe},
volume = {6},
number = {8},
pages = {105},
year = {2020}
}

@article{Kolobov2021,
author={Kolobov, V. I. and Golubkov, K. and Muñoz de Nova, J. R. and Steinhauer, J.},
title={Observation of stationary spontaneous {Hawking} radiation and the time evolution of an analogue black hole},
journal={Nat. Phys},
volume={17}, 
pages={362–367},
year={2021},
doi={10.1038/s41567-020-01076-0}
}

@article{Fabbri2021,
  title = {Ramp-up of {Hawking} Radiation in {Bose-Einstein-Condensate} Analog Black Holes},
  author = {Fabbri, Alessandro and Balbinot, Roberto},
  journal = {Phys. Rev. Lett.},
  volume = {126},
  issue = {11},
  pages = {111301},
  numpages = {6},
  year = {2021},
  month = {Mar},
  publisher = {American Physical Society},
  doi = {10.1103/PhysRevLett.126.111301},
  url = {https://link.aps.org/doi/10.1103/PhysRevLett.126.111301}
}

@article{Micheli2022,
  title = {Phonon decay in one-dimensional atomic {Bose} quasicondensates via {Beliaev-Landau} damping},
  author = {Micheli, Amaury and Robertson, Scott},
  journal = {Phys. Rev. B},
  volume = {106},
  issue = {21},
  pages = {214528},
  numpages = {23},
  year = {2022},
  month = {Dec},
  publisher = {American Physical Society},
  doi = {10.1103/PhysRevB.106.214528},
  url = {https://link.aps.org/doi/10.1103/PhysRevB.106.214528}
}

@article{Baak2022,
  title = {Number-conserving solution for dynamical quantum backreaction in a {Bose-Einstein} condensate},
  author = {Baak, Sang-Shin and Ribeiro, Caio C. Holanda and Fischer, Uwe R.},
  journal = {Phys. Rev. A},
  volume = {106},
  issue = {5},
  pages = {053319},
  numpages = {14},
  year = {2022},
  month = {Nov},
  publisher = {American Physical Society},
  doi = {10.1103/PhysRevA.106.053319},
  url = {https://link.aps.org/doi/10.1103/PhysRevA.106.053319}
}

@article{Duval2023,
  title = {Quantum kinetics of quenched two-dimensional {Bose} superfluids},
  author = {Duval, Cl\'ement and Cherroret, Nicolas},
  journal = {Phys. Rev. A},
  volume = {107},
  issue = {4},
  pages = {043305},
  numpages = {13},
  year = {2023},
  month = {Apr},
  publisher = {American Physical Society},
  doi = {10.1103/PhysRevA.107.043305},
  url = {https://link.aps.org/doi/10.1103/PhysRevA.107.043305}
}

@article{Butera2023,
  title = {Numerical Studies of Back Reaction Effects in an Analog Model of Cosmological Preheating},
  author = {Butera, Salvatore and Carusotto, Iacopo},
  journal = {Phys. Rev. Lett.},
  volume = {130},
  issue = {24},
  pages = {241501},
  numpages = {7},
  year = {2023},
  month = {Jun},
  publisher = {American Physical Society},
  doi = {10.1103/PhysRevLett.130.241501},
  url = {https://link.aps.org/doi/10.1103/PhysRevLett.130.241501}
}

@article{Pal2024,
  title = {Quantum nonlinear effects in the number-conserving analog gravity of {Bose-Einstein} condensates},
  author = {Pal, Kunal and Fischer, Uwe R.},
  journal = {Phys. Rev. D},
  volume = {110},
  issue = {11},
  pages = {116022},
  numpages = {15},
  year = {2024},
  month = {Dec},
  publisher = {American Physical Society},
  doi = {10.1103/PhysRevD.110.116022},
  url = {https://link.aps.org/doi/10.1103/PhysRevD.110.116022}
}

@article{Micheli2024,
     author = {Micheli, Amaury and Robertson, Scott},
     title = {Dissipative parametric resonance in a modulated {1D} {Bose} gas},
     journal = {Comptes Rendus. Physique},
     publisher = {Acad\'emie des sciences, Paris},
     year = {2024},
     volume={25},
     pages={1--34},
     doi = {10.5802/crphys.250},
}

@phdthesis{Ciliberto2024_PhD,
author       = {Ciliberto, Giorgio}, 
title        = {Non-locality and back-reaction in acoustic black holes
and non-linearity in quantum fluid dynamics},
school = {{Universit{\'e} Paris-Saclay and Albert-Ludwigs-Universit{\"a}t Freiburg}},
year         = 2024,
month        = 12,
url = {https://theses.hal.science/tel-04889022}
}

@article{delRio2025,
author={del Rio, A.},
title={The backreaction problem for black holes in semiclassical gravity},
journal={Gen. Relativ. Gravit.},
volume={57}, 
pages={30},
year={2025},
doi={10.1007/s10714-025-03352-x}
}

\end{document}